\pdfoutput=1
\documentclass{aa}

\usepackage{amsmath,amssymb}
\usepackage[varg]{txfonts}

\usepackage{ulem}
\usepackage{natbib}

\usepackage{color}
\usepackage{graphicx}

\usepackage{epstopdf}
\usepackage{gensymb}
\usepackage{stfloats}

\usepackage[utf8]{inputenc}
\usepackage{hyperref}
\usepackage{booktabs,multirow}

\definecolor{mygray}{gray}{0.6}
\definecolor{darkblue}{rgb}{0.0, 0.0, 0.8}

\newcommand{\se}[1]{Sect.~\ref{sec:#1}}
\newcommand{\eq}[1]{Eq.~(\ref{eq:#1})}

\newcommand{\fg}[1]{Fig.~\ref{fig:#1}}
\newcommand{\Fg}[1]{Figure~\ref{fig:#1}}
\newcommand{\tb}[1]{Table~\ref{tab:#1}}

\begin{document}

\title{A Lagrangian Model for Dust Evolution in Protoplanetary Disks:\\ Formation of Wet and Dry Planetesimals at Different Stellar Masses}
\titlerunning{A Lagrangian Model for Dust Evolution in Protoplanetary Disks}
\author{Djoeke Schoonenberg\inst{1}, Chris W.~Ormel\inst{1}, Sebastiaan Krijt\inst{2, 3}}
\authorrunning{D.~Schoonenberg et al.}
\institute{Anton Pannekoek Institute for Astronomy, University of Amsterdam, Science Park 904, 1090 GE Amsterdam, The Netherlands\label{inst1} \\
\email{d.schoonenberg@uva.nl}
\and
Department of the Geophysical Sciences, The University of Chicago, 5734 South Ellis Avenue, Chicago, IL 60637, USA\label{inst2}
\and
Hubble Fellow\label{inst3}}
\date{\today}

\abstract{We introduce a new Lagrangian smooth-particle method to model the growth and drift of pebbles in protoplanetary disks. The Lagrangian nature of the model makes it especially suited to follow characteristics of individual (groups of) particles, such as their composition. In this work we focus on the water content of solid particles. Planetesimal formation via streaming instability is taken into account, partly based on previous results on streaming instability outside the water snowline that were presented in \citet{SO2017}. We validate our model by reproducing earlier results from the literature and apply our model to steady-state viscous gas disks (with constant gas accretion rate) around stars with different masses. We also present various other models where we explore the effects of pebble accretion, the fragmentation velocity threshold, the global metallicity of the disk, and a time-dependent gas accretion rate.

We find that planetesimals preferentially form in a local annulus outside the water snowline, at early times in the lifetime of the disk ($\lesssim$$10^{5} \: \rm{yr}$), when the pebble mass fluxes are high enough to trigger the streaming instability. During this first phase in the planet formation process, the snowline location hardly changes due to slow viscous evolution, and we conclude that assuming a constant gas accretion rate is justified in this first stage.

The efficiency of converting the solids reservoir of the disk to planetesimals depends on the location of the water snowline. Cooler disks with a closer-in water snowline are more efficient at producing planetesimals than hotter disks where the water snowline is located further away from the star. Therefore, low-mass stars tend to form planetesimals more efficiently, but any correlation may be overshadowed by the spread in disk properties.
}

\maketitle
\section{Introduction}
Rocky planets and the cores of gas giants form from micron-sized dust grains in gaseous disks around young stars. It is generally accepted that an intermediate stage on the way from small dust grains in protoplanetary disks to full-sized planets is the formation of $\sim$kilometer-sized planetesimals, which mark the transition to a gravitation-dominated growth phase \citep{Safronov1969,PollackEtal1996,2000SSRv...92..279B}. Planetesimal formation is still an active field of research since theories face several problems. First, typical relative velocities between $\sim$cm-sized particles are often too large for coagulation, such that particles fragment or bounce off each other upon collision, rather than stick to form even larger particles \citep{2000Icar..143..138B,2010A&A...513A..57Z}. Second, particles that are large enough to be aerodynamically decoupled from the gas disk (`pebbles') lose angular momentum and drift towards the central star \citep{Whipple1972,Weidenschilling1977}: planetesimals need to form before the solids are lost due to this process.
 
 A promising method to form planetesimals directly from small particles -- without the need for growth through all intermediate sizes -- is by streaming instability: drifting pebbles clump together and can collapse under their own gravity to form planetesimals \citep{2005ApJ...620..459Y,2007Natur.448.1022J,2007ApJ...662..627J}. For streaming instability to operate, however, a solids-to-gas ratio that is enhanced with respect to the canonical value of 1\% is required \citep{2009ApJ...704L..75J,BaiStone2010i,2015A&A...579A..43C,2017A&A...606A..80Y}. In classical planet formation theory, planetesimals are assumed to form throughout the entire disk, but various studies propose that an enhanced solids-to-gas ratio forms preferentially at specific locations instead, {\it e.g.} in the inner disk \citep{2016A&A...594A.105D}; in the outer disk at late times \citep{2017ApJ...839...16C}; or near the water snowline, where water transitions from the vapour phase to the solid phase \citep{2011ApJ...728...20S,2013A&A...552A.137R,2016A&A...596L...3I,ArmitageEtal2016,SO2017,2017A&A...608A..92D}. Detections by the Atacama Large Millimeter/submillimeter Array (ALMA) of substructures in protoplanetary disks such as axisymmetric gaps and rings \citep{2015ApJ...808L...3A,2015ApJ...802L..17A,2016ApJ...819L...7N,2016PhRvL.117y1101I,2018A&A...610A..24F} may support the view that planetesimals form at specific distances from the star. 
\citet{OLS2017} presented a complete formation scenario for the TRAPPIST-1 planets that is based on the formation of planetesimals at a single location: the water snowline. However, this scenario still lacks a thorough numerical model.

In this work we present a new, versatile model for planetesimal formation. We follow the growth and drift of pebbles in protoplanetary disks and include planetesimal formation via the streaming instability, partly based on the results of \citet{SO2017} (hereafter: SO17). In SO17 we presented a local model where we investigated whether water diffusion and condensation could lead to conditions conducive to streaming instability outside the water snowline, and how this depends on certain parameters such as the turbulence strength and the particle size. In contrast to SO17, in the current work we consider the entire protoplanetary disk. The pebble mass flux and particle sizes are not treated as input parameters as in SO17, but follow self-consistently from the simulation and evolve in time. Another difference with SO17 is that in this paper, the planetesimal formation process is followed (mass is removed from pebbles as planetesimals form), whereas in SO17 we focused only on the conditions for planetesimal formation.
Our model makes use of the Lagrangian method where super-particles represent groups of particles with identical physical properties. In the context of dust evolution in protoplanetary disks, this method was pioneered by several studies ({\it e.g.} \citet{2008A&A...487..265L,2016A&A...586A..20K,2017MNRAS.467.1984G}). In contrast to Eulerian approaches to dust evolution ({\it e.g.} \citet{2012A&A...539A.148B,2012ApJ...752..106O}), this method is mesh-free: physical quantities are computed at the locations of the super-particles, not at the locations of grid cells, and individual particle characteristics such as water content can easily be followed as particles move through the disk.

Exoplanet data shows that low-mass stars are efficient at forming super-Earths \citep{2015ApJ...814..130M,2018arXiv180500023M}. Differences in the occurrence rates between stars of different masses could (partly) originate in differences in planetesimal formation efficiencies. In this paper we apply our model to testing the efficiency of icy and dry planetesimal formation as a function of stellar mass. We find that generally, low-mass stars convert a larger fraction of the solids reservoir in their disks to planetesimals than high-mass stars, but other disk parameters play important roles as well. 
We also discuss the effects that pebble accretion, the viscous evolution of the gas disk, changing the metallicity and fragmentation threshold have on planetesimal formation.

We describe the different components of our model in Sect.~2--4 and provide a summary of the model in Sect.~5. The results are discussed in Sect.~6--7 and we discuss our findings in Sect.~8. Our main conclusions are listed in Sect.~9.

\section{Gas Disk Model}\label{sec:gas}
\subsection{Surface density profile}
In our standard disk model, we adopt for simplicity a steady-state gas surface density profile $\Sigma_{\rm{gas}}$ for a given gas accretion rate $\dot{M}_{\rm{gas}}$ and dimensionless turbulence parameter $\alpha$ \citep{1973A&A....24..337S,1974MNRAS.168..603L}:
\begin{equation}\label{eq:svis}
\Sigma_{\rm{gas}} = \frac{\dot{M}_{\rm{gas}}}{3 \pi \nu}
\end{equation}
where the viscosity $\nu$ is related to $\alpha$ as follows:
\begin{equation}\label{eq:nu}
\nu = \alpha c_{s}^{2} \Omega^{-1}
\end{equation}
with $c_{s}$ the sound speed and $\Omega$ the Keplerian orbital frequency. The gas moves inward at a speed $|v_{\rm{gas}}| = 3 \nu / 2r$ where $r$ is the radial distance from the star. The surface density $\Sigma_{\rm{gas}}$ for our fiducial model (\tb{inputpar}) is plotted by the black line in \fg{gasdisk}.

We will discuss the validity of the assumption of constant $\dot{M}_{\rm{gas}}$ in the context of our model, and investigate the effects of relaxing this assumption in \se{mdottime}.

\subsection{Temperature profile}\label{sec:temp}
We consider two mechanisms that heat the protoplanetary disk: viscous heating and stellar irradiation. Viscous heating is only important in the innermost region of the disk and leads to a radial dependence of $T_{\rm{visc}} \propto r^{-3/4}$ \citep{2002apa..book.....F}. Stellar irradiation results in a temperature profile in the outer disk that goes as $T_{\rm{irr}} \propto r^{-1/2}$ \citep{1987ApJ...323..714K}. These general temperature profiles can only be directly used for the disk midplane temperature (of interest in this work) if the vertical optical depth were radially constant, which is probably not the case. A more sophisticated model would calculate the temperature from the dust properties self-consistently. However, for the purposes of the current work the simple power laws suffice. 
For our fiducial model, we use the following viscous and irradiated midplane temperature profiles:
\begin{equation}
T_{\rm{visc}} (r) = 350 \left(\frac{r}{1 \: \rm{au}}\right)^{-3/4}
\end{equation}
\begin{equation}
T_{\rm{irr}} (r) = 177 \left(\frac{r}{1 \: \rm{au}}\right)^{-1/2}
\end{equation}
We take the global temperature profile $T (r)$ to be the (smoothed) maximum of the temperature profiles $T_{\rm{visc}}$ and $T_{\rm{irr}}$:
\begin{equation}\label{eq:temp}
T (r) = [T_{\rm{visc}}^{4}(r) + T_{\rm{irr}}^{4}(r)]^{1/4}
\end{equation}
which then defines the isothermal sound speed profile $c_{s}$:
\begin{equation}
c_{s} (r) = \sqrt{\frac{k_{B} T (r)}{\mu}}  
\end{equation}
where $k_{B}$ is the Boltzmann constant and $\mu$ is the mean molecular weight of the gas, for which we take a value of 2.34 times the proton mass appropriate for a solar metallicity gas. The temperature profiles $T$, $T_{\rm{visc}}$, and $T_{\rm{irr}}$ for our fiducial model (\tb{inputpar}) are plotted by the blue lines in \fg{gasdisk}.

We assume that the disk is vertically isothermal, such that the disk scale height $H_{\rm{gas}}$ is given by:
\begin{equation}
H_{\rm{gas}} = c_{s} / \Omega
\end{equation}

\begin{figure}[t]
	\centering
		\includegraphics[width=0.49\textwidth]{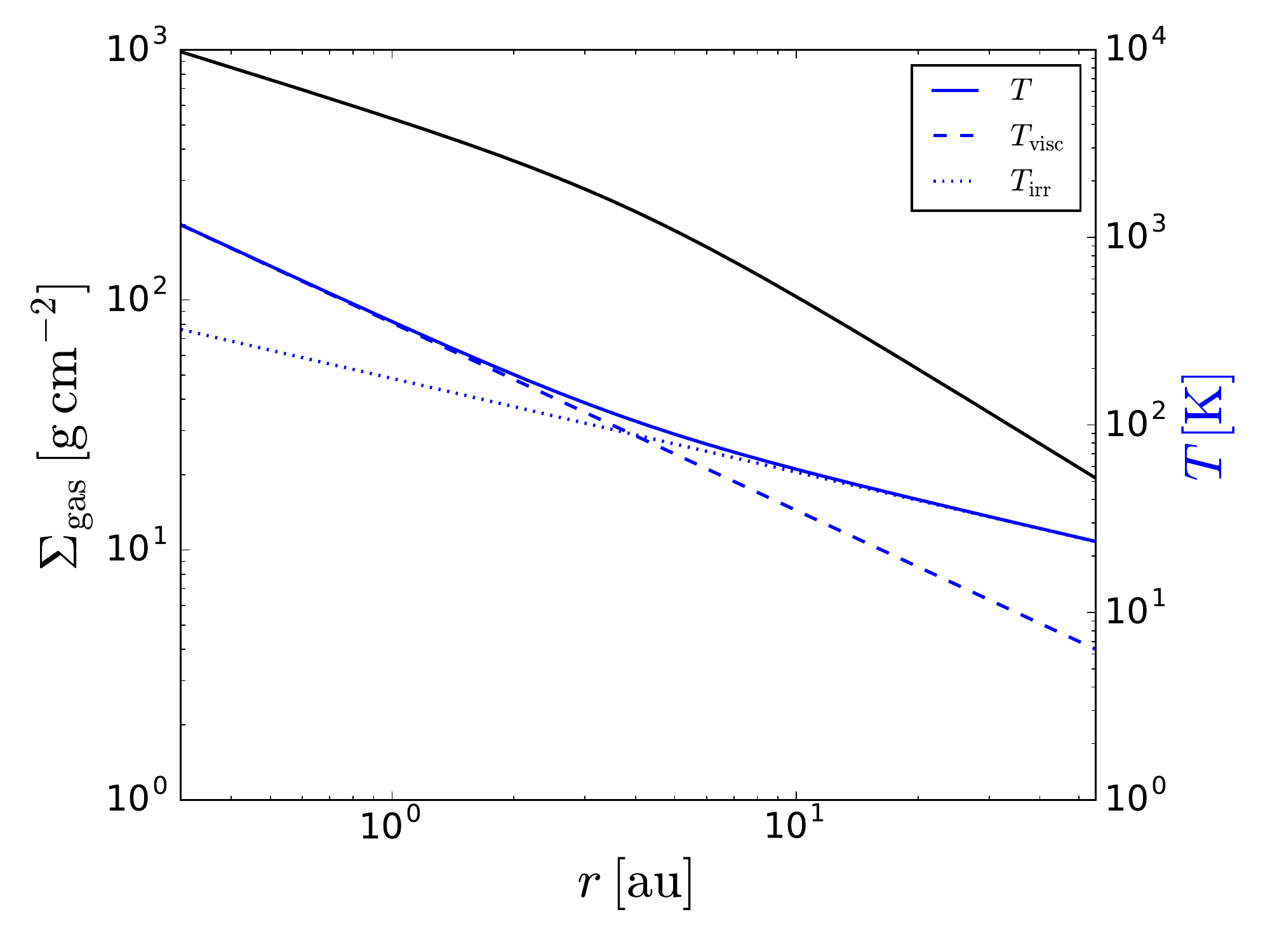}
\caption{Gas disk profiles for our fiducial model (\tb{inputpar}). The black line corresponds to the gas surface density profile $\Sigma_{\rm{gas}}$ (\eq{svis}); the solid blue line corresponds to the temperature profile $T$, which receives contributions from viscous heating (dashed blue line) and stellar irradiation (dotted blue line) (\eq{temp}).\label{fig:gasdisk}}
\end{figure}

\section{Treatment of Solids: Lagrangian Smooth-Particle Model}\label{sec:solids}
In this work we adopt a Lagrangian method to solve for the growth and the radial movement of the solid particles in the disk. A clear advantage of the Lagrangian method is that it is naturally suited to follow characteristics of individual particles, such as their composition and porosity, as they grow and move through the disk. It is not possible to follow all particles in the disk, however. We therefore use super-particles: groups of particles with the same physical properties. In this work we consider two different classes of super-particles: dust/pebble super-particles\footnote{We use the term `dust' for solid particles that are well-coupled to the gas, and the term `pebbles' for particles that have a non-negligible radial drift velocity (Stokes numbers of $\sim$$10^{-3} - 10^{1}$; see \se{grdr}). Throughout the paper, the term `pebble super-particle' is used instead of `dust/pebble super-particle'.}, and planetesimal super-particles (in future work, more super-particle classes may be added to the model). Pebble super-particles can move through the disk, whereas planetesimal super-particles are inert. We assume a mono-disperse particle size distribution at each point in space and time, such that a pebble super-particle represents particles that all have the same individual particle mass.

The characteristics of the pebble super-particles that we follow are:
\begin{itemize}
\item location
\item total mass
\item individual particle mass
\item water mass fraction
\end{itemize}
The planetesimal super-particles in our model are characterised by:
\begin{itemize}
\item location (fixed)
\item total mass
\item water mass fraction
\end{itemize}
The total mass of a planetesimal super-particle can change because of streaming instability and pebble accretion: when planetesimal formation and/or pebble accretion occurs, mass is transferred from the pebble super-particles to the planetesimal super-particles. 

Formally, we end up with a system of ordinary differential equations (ODEs) that govern the time evolution of the super-particles, of the form:
\begin{equation}\label{eq:ode}
\frac{d X_{i, j}}{d t} = \vec{Y} (\vec{X})
\end{equation}
where $i$ indicates the super-particle number, and $j$ corresponds to their properties: radial position $r$, individual particle mass $m_{p}$ (for pebble super-particles only), total super-particle mass $M$. The right-hand side of \eq{ode} is a vector $\vec{Y}$ that consists of the planetesimal formation rates, drift velocities, and particle growth rates, which depend on the properties of the super-particles $\vec{X}$. We solve \eq{ode} making use of the  \texttt{Python} package \texttt{scipy.integrate.ode}.

\subsection{Smooth-particle method}
In \citet{2016A&A...586A..20K}, the solids surface density and its radial derivative at the location of each super-particle is calculated using a `tripod' method: each super-particle consists of three `legs' that can move closer to each other (higher surface density) or further apart (lower surface density). A consequence of this approach is that the super-particles evolve independently of each other. In contrast, in our method all super-particles are connected. As in SPH methods (e.g. \citet{2008A&A...487..265L}), we approximate the solids surface density at each super-particle location using a weighting kernel $W$, that accounts for the contribution of neighbouring super-particles as function of their mass and distance. Therefore, we simultaneously solve for the evolution of all solids in the disk. As in \citet{2016A&A...586A..20K} but unlike \citet{2008A&A...487..265L}, our model is one-dimensional; we only deal with the radial dimension $r$. The vertical dimension of the disk is taken into account by means of the gas and solids scale heights, and the model is symmetric in the azimuthal dimension.

The value of a quantity $F$ at each particle location $x$ is kernel-approximated by:
\begin{equation}\label{eq:smooth}
F(x) = \int_{\Omega} F(x') W(x-x', h) dx'
\end{equation}
where $x$ and $x'$ are vectors and $W(x-x', h)$ is the kernel, for which we take \citep{hicks, liu2003smoothed}:
\begin{equation}\label{eq:kernel}
W(\Delta x, h) \equiv \max[0, \frac{3}{4 h} (1 - [\Delta x / h]^{2})]
\end{equation}
where $h$ is the smoothing length, $\Delta x = |x - x'|$ is the absolute distance between the location of super-particle $x'$ and the location of the super-particle-of-interest $x$. The prefactor ensures that the kernel is normalised. Particles that are separated from $x$ by a distance larger than the smoothing length ($|x - x'| > h$) do not influence the value $F(x)$; in other words, the kernel $W$ is compact.

Since the values of $F(x)$ are only defined at discrete locations (the super-particle locations), we turn the integral in \eq{smooth} into a sum over all simulated super-particles. For the surface density $\Sigma$ at the location $x_{i}$ of super-particle $i$, we then find:
\begin{equation}
\Sigma (x_{i}) = \frac{1}{2 \pi x_{i}} \sum_{j \in {\rm{support}}} M_{j} W(|x_{i} - x_{j}|, h_{i})
\end{equation}
where $M_{j}$ is the mass of supporting super-particle $j$ and $h_{i}$ is the smoothing length of super-particle $i$. We treat the smoothing length $h_i$ as a variable, demanding that at each super-particle location and time $h_i$ takes on a value such that there are five neighbouring super-particles (including super-particle $i$ itself) in the support group contributing to the density at location $x_i$ (for the simulations presented in this paper, changing the number of neighbours to three or seven did not change the results). The advantage of a variable smoothing length is that the code can adapt to regions of high density as well as regions of low density. Because of the compactness of the kernel $W$, for each super-particle we only have to sum over the super-particles in its support group.

\subsection{Boundary treatment and initial conditions}\label{sec:bound}
The condition we set on the inner boundary of the simulated disk is a constant solids surface density gradient. The innermost particles --- which lack particles interior to them to fill their support group --- get assigned a surface density value in agreement with the surface density gradient in the inner region.  

\noindent
At the outer boundary, we use an exponentially cut-off initial solid surface density profile, such that particles close to the outer boundary barely grow and drift over the lifetime of the disk, to prevent any unwanted outer boundary effect.

At the start of the disk evolution, a fraction of the total disk mass is in the form of dust grains (ice + silicates). We take this fraction, also called the metallicity $Z$, to be initially constant throughout the disk with a sharp exponential cutoff at the outer disk radius $r_{\rm{out}}$. Therefore, the solids surface density profile $\Sigma_{\rm{solids}}$ initially follows:
\begin{equation}
\Sigma_{\rm{solids}} (r, t = 0) = Z \Sigma_{\rm{gas}} (r) \exp{[-(r/r_{\rm{out}})^{4}]}
\end{equation}
where the sharp exponential cut-off is just to ensure that no unwanted numerical effects occur at the outer boundary.
The initial size of solid particles is set to 0.1 $\mu m$. We assume that in the outer disk, the water ice mass fraction $f_{\rm{ice}}$ of particles is 50\% ($f_{\rm{ice, out}} = 0.5$) \citep{2003ApJ...591.1220L,2015Icar..258..418M}. Interior to the water snowline, the solids consist of pure silicate ($f_{\rm{ice, in}} = 0$). The location of the water snowline and the transition between $f_{\rm{ice, out}}$ and $f_{\rm{ice, in}}$ will be discussed in \se{snow}.

The initial mass of a super-particle follows from the initial solids surface density profile $\Sigma_{\rm{solids}} (r)$ and the initial number of super-particles $N$. We discretise the radial distance to the central star $r$ into $N$ annuli with edges $r_{i}$ between the inner edge of the disk $r_{\rm{in}}$ and the outer edge $r_{\rm{out}}$. The $k$th super-particle then gets assigned a mass $m_{k} = \int_{r_{k}}^{r_{k+1}} 2 \pi r \Sigma (r) dr$ (corresponding to the total mass in the $k$th annulus) and an initial position $x_{k} = (r_{k+1} - r_{k}) / 2$.
If the initial solids surface density profile is proportional to $r^{-1}$ ($\Sigma_{\rm{solids}} (r) = \Sigma_{0} r^{-1}$ where $\Sigma_{0}$ is constant) and the spacing between the annuli is linear, each super-particle has the same initial mass $m = 2 \pi \Delta r \Sigma_{0}$ with $\Delta r$ the annulus width. Alternatively, the initial particle locations can be chosen such that the resolution varies across the disk (for example, closely-spaced particles in the inner disk and particles that are further apart in the outer disk); in that case the initial super-particle masses differ. In this work we use a logarithmic spacing between the initial positions of the dust/pebble super-particles, so that super-particles in the outer disk have larger total masses than super-particles in the inner disk.

\subsection{Resampling}
During the evolution of the disk, the spacing between adjacent super-particles can become larger than desired. This can for example occur when the resolution needs to be high in the inner disk or around a special location ({\it e.g.}, the snowline), whereas in the outer disk the particles are initially further apart. Due to radial drift, closely-spaced particles from the inner disk get accreted to the star and less closely-spaced particles from the outer disk enter the inner disk, such that the resolution in the inner disk decreases.
If at any point the resolution becomes too low, we initiate a `resampling' algorithm. In our model we resample when the separation between particles at a certain point becomes 20\% larger than the initial particle separation at that point.
The number and locations of super-particles are reset to the initial configuration, and the characteristics of the new super-particle population are sampled from the super-particle population right before the resampling process. The total mass of each new super-particle is extracted from the cumulative mass distribution right before `resampling', in order to ensure mass conservation. \Fg{resample} shows a cartoon of the concept of resampling.

\begin{figure}[t]
	\centering
		\includegraphics[width=0.5\textwidth]{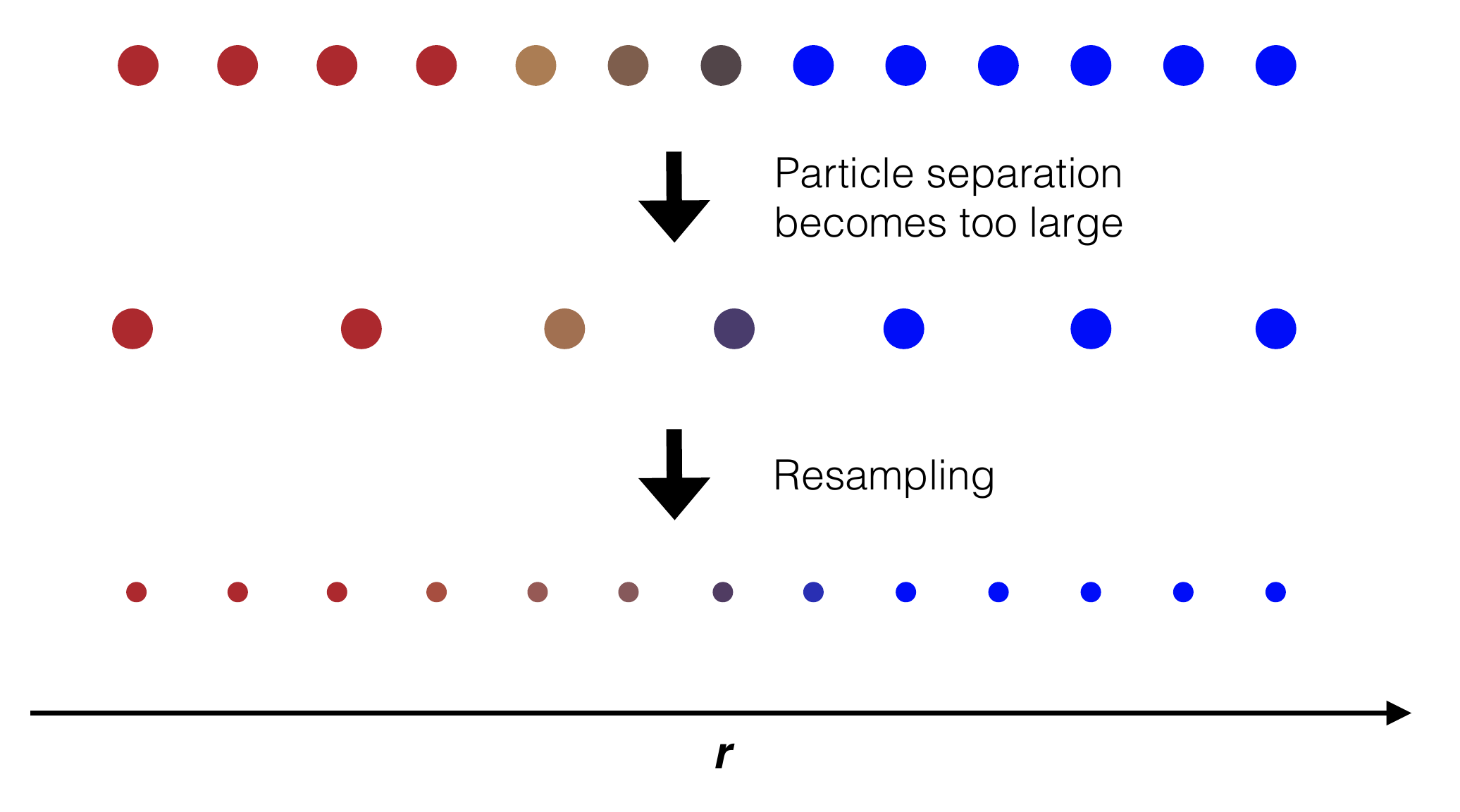}
\caption{Procedure cartoon of the resampling concept. When the separation between neighbouring pebble super-particles becomes too large, the number and configuration of particles is reset such that the resolution is again equal to that of the initial set-up. The total mass in pebble super-particles is conserved and the characteristics (such as water ice fraction, indicated by colour here) are sampled from the situation before resampling.\label{fig:resample}}
\end{figure}

\subsection{Particle growth and radial drift}\label{sec:grdr}
In this section we describe how we treat dust evolution (particle growth and radial drift) in our model. In Appendix~A we test our model against results from the literature.

The dust population at the locations of the super-particles is represented by a mono-disperse dust size distribution \citep{2016A&A...586A..20K,SatoEtal2016}. This means that the dust particles represented by one super-particle all have the same size $s_{p}$ and mass $m_{p}$. Within each super-particle then, particles can coagulate, increasing the individual particle mass $m_{p}$ of that super-particle. The mass growth rate $dm_{p} / dt$ assuming perfect sticking is \citep{2016A&A...586A..20K}:
\begin{equation}\label{eq:massgrowthrate}
\frac{d m_{p}}{d t} = \frac{\Sigma_{\rm{solids}}}{\sqrt{2 \pi} H_{\rm{solids}}} \sigma_{\rm{col}} v_{\rm{rel}}
\end{equation}
with $H_{\rm{solids}}$ the solids scale height, $\sigma_{\rm{col}}$ the collisional cross-section (which we take equal to the geometrical cross-section), and $v_{\rm{rel}}$ the relative velocity between particles. As long as the particles are small and well-coupled to the gas, $H_{\rm{solids}} = H_{\rm{gas}}$. Once particles have grown to sizes where they start to aerodynamically decouple from the gas disk, their scale height is reduced with respect to that of the gas due to vertical settling, and is given by \citep{2007Icar..192..588Y}:
\begin{equation}\label{eq:hsolids}
H_{\rm{solids}} = H_{\rm{gas}} \sqrt{\frac{\alpha}{\tau + \alpha}}
\end{equation}
where $\tau$ is the dimensionless stopping time, which measures the degree of coupling between the particle and the gas. We distinguish between two regimes of $\tau$: in the Stokes regime, the particle size is larger than the mean-free path of the gas molecules $l_{\rm{mfp}}$, and the stopping time is calculated in a fluid description; in the Epstein regime, the particle size is smaller than the mean-free path of the gas molecules, and a particle description is required instead \citep{Weidenschilling1977}. The mean-free path $l_{\rm{mfp}}$ is given by:
\begin{equation}\label{eq:lmfp}
l_{\rm{mfp}} = \frac{\mu}{\sqrt{2} \rho_{\rm{gas}} \sigma_{\rm{mol}}}
\end{equation}
where $\rho_{\rm{gas}}$ is the gas volume density at the disk midplane ($\rho_{\rm{gas}} = \Sigma_{\rm{gas}} / \sqrt{2 \pi} H_{\rm{gas}}$) and $\sigma_{\rm{mol}}$ is the molecular collisional cross-section, for which we take $\sigma_{\rm{mol}} = 2 \times 10^{-15} \: \rm{cm}^{2}$ as appropriate for hydrogen \citep{chapman1970mathematical}.
The dimensionless stopping time is given by:
\begin{equation}\label{eq:stoppingtime}
\tau = \begin{cases} \displaystyle \Omega \frac{\rho_{\bullet,p} s_{p}}{v_{\rm{th}} \rho_{\rm{gas}}}, &\text{(Epstein:}\: \: \: s_{p} < \frac{9}{4} l_{\rm{mfp}})\\[1em]
\displaystyle \Omega \frac{4 \rho_{\bullet,p} s_{p}^{2}}{9  l_{\rm{mfp}} v_{\rm{th}} \rho_{\rm{gas}}}, &\text{(Stokes:}\: \: \: s_{p} > \frac{9}{4} l_{\rm{mfp}})
\end{cases}
\end{equation}
where $s_{p}$ is the particle radius, $\rho_{\bullet,p}$ is the particle internal density, and $v_{\rm{th}}$ is the thermal velocity of the gas molecules, defined as $v_{\rm{th}} = \sqrt{8/\pi} c_{s}$. 
The particle radius and internal density are determined from the particle mass $m_{p}$ and water mass fraction $f_{\rm{ice}}$: 
\begin{align}
\rho_{\bullet,p} &= \frac{\rho_{\bullet,\rm{ice}} \rho_{\bullet,\rm{sil}} m_{p}}{(1 - f_{\rm{ice}}) m_{p} \rho_{\bullet,\rm{ice}} + f_{\rm{ice}} m_{p} \rho_{\bullet,\rm{sil}}}\\
s_{p} &=  \left(\frac{3 m_{p}}{4 \pi \rho_{\bullet, p}}\right)^{1/3}
\end{align}
where $\rho_{\bullet,\rm{ice}} = 1 \: \rm{g} \: \rm{cm}^{-3}$ is the density of a pure ice particle and $\rho_{\bullet,\rm{sil}} = 3 \: \rm{g} \: \rm{cm}^{-3}$ the density of a pure silicate particle.

For the relative velocity between individual particles within a super-particle $v_{\rm{rel}}$, we consider the relative velocity due to turbulence, Brownian motion, and radial and azimuthal drift, in the same way as \citet{2016A&A...586A..20K}. The total relative velocity is given by the maximum of these four contributions. We find that in most cases, the turbulent relative velocity $v_{\rm{rel, turb}}$ dominates.
In the so-called intermediate regime --- where the turn-over time scale of the smallest eddies ($t_{\eta} = \rm{Re}_{\rm{T}}^{-1/2} \Omega^{-1}$ where $\rm{Re}_{\rm{T}}$ is the turbulence Reynolds number) is much smaller than the stopping time of the particles \citep{2007A&A...466..413O}, which is the regime of importance in our model --- $v_{\rm{rel, turb}}$ is given by:
\begin{equation}
v_{\rm{rel, turb}} \sim \sqrt{3 \alpha \tau} c_{s}
\end{equation}
Therefore, the larger the particles grow, the more violently they impact each other. 

\subsubsection{Fragmentation}\label{sec:frag}
When the relative velocity $v_{\rm{rel}}$ between particles becomes larger than a certain fragmentation threshold $v_{\rm{frag}}$, particles start to fragment rather than coagulate when they collide. Concerning $v_{\rm{frag}}$ we distinguish between icy particles and dry particles. The surface energy of an aggregate containing water ice is larger than that of a silicate one, and therefore icy particles are more sticky than silicate particles \citep{1993ApJ...407..806C}. Motivated by laboratory experiments, for the fragmentation threshold velocity of silicate particles we take $v_{\rm{frag, sil}}$ = $3 \: \rm{m} \: \rm{s}^{-1}$ \citep{1993Icar..106..151B}; and for icy particles we take $v_{\rm{frag, ice}}$ = $10 \: \rm{m} \: \rm{s}^{-1}$ \citep{1999A&A...347..720S,2015ApJ...798...34G}. The fragmentation threshold $v_{\rm{frag}}$ for particles with ice fraction $f_{\rm{ice}}$ is then given by:
\begin{equation}\label{eq:fragvel}
v_{\rm{frag}} (f_{\rm{ice}}) = \frac{f_{\rm{ice}}}{f_{\rm{ice, out}}} v_{\rm{frag, ice}} + \left(1 - \frac{f_{\rm{ice}}}{f_{\rm{ice, out}}}\right) v_{\rm{frag, sil}}
\end{equation}
where we simply coupled the two fragmentation velocities according to the ice mass fraction $f_{\rm{ice}}$ (similar to \citet{2016A&A...587A.128L}, who interpolate the fragmentation velocities using the fractional abundance of icy monomers). Our assumption that the fragmentation velocity increases linearly with $f_{\rm{ice}}$ is arbitrary but unimportant, since $f_{\rm{ice}}$ changes rapidly across the water snowline \citep{2017A&A...608A..92D}.

If $v_{\rm{rel}} >  v_{\rm{frag}}$, we instantaneously set the particle mass $m_{p}$ to the fragmentation mass limit $m_{\rm{frag}}$. In \fg{mfrag} we have plotted $m_{\rm{frag}}$ as a function of semi-major axis $r$ for our fiducial model.
The assumption that fragmentation is instantaneous is justified by the fact that $m_{\rm{frag}}$ increases for particles that are drifting from the outer parts of the disk inward within the drift-limited region. In this region, drift is faster than growth/fragmentation \citep{2012A&A...539A.148B}). In the viscosity-dominated inner disk, $m_{\rm{frag}}$ decreases for inward-drifting particles (see \fg{mfrag}). In this region therefore, inward-drifting particles that have a mass that is limited to $m_{\rm{frag}}$ are constantly fragmenting to smaller masses on their way to the star. However, the viscosity-dominated region is fragmentation-limited, meaning that growth/fragmentation are faster than drift \citep{2012A&A...539A.148B}. Therefore, as long as the region where the temperature is dominated by viscous heating is in the fragmentation-limited regime, treating fragmentation as an instantaneous process is justified throughout the disk.

\begin{figure}[t]
	\centering
		\includegraphics[width=0.5\textwidth]{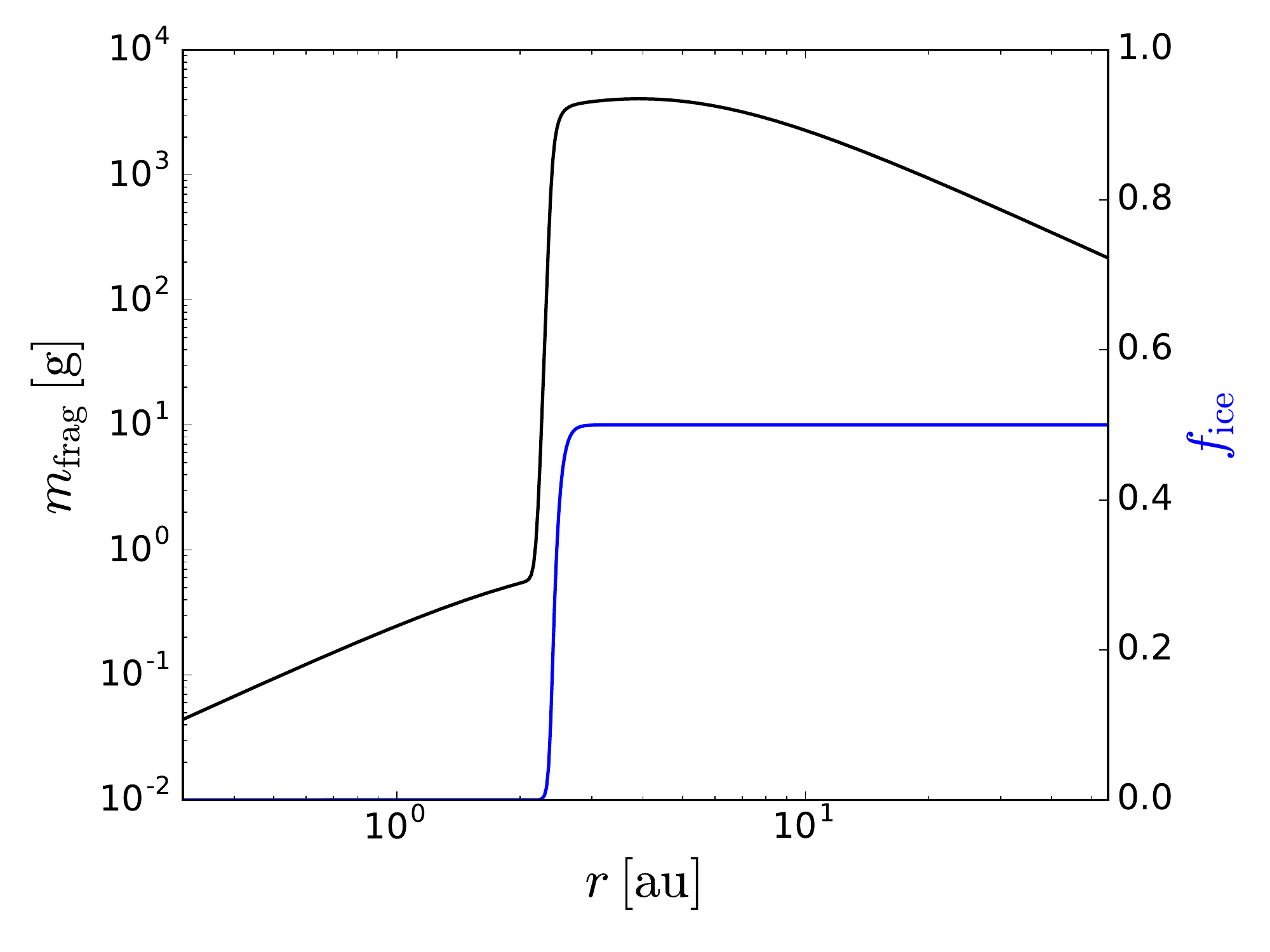}
\caption{Fragmentation mass limit $m_{\rm{frag}}$ (black line) for our fiducial model with initial ice mass fraction profile $f_{\rm{ice}}$ (blue line). The snowline is located at around 2.4 au. Because of the different stickiness between icy and silicate particles, the fragmentation velocity, and therefore the fragmentation mass limit, varies rapidly across the snowline.\label{fig:mfrag}}
\end{figure}

\subsubsection{Radial drift}
The radial velocity of solid particles $v_{p}$ (taking into account the back-reaction of the solids onto the gas) is given by \citep{NakagawaEtal1986}:
\begin{equation}\label{eq:vdriftnoce}
v_{p} = - \frac{2 \eta v_K \tau - v_{\rm{gas}} (1 + \xi)}{(1 + \xi)^{2} + \tau^{2}}
\end{equation}
where $v_{\rm{gas}}$ is negative (the gas is moving inward as well), $\xi$ is the midplane solids-to-gas volume density ratio, and $\eta v_K$ is the difference between the azimuthal motion of the gas disk and the Keplerian velocity $v_K$:
\begin{equation}
    \label{eq:eta}
\eta v_{K} = -\frac{1}{2} \frac{c_{s}^{2}}{v_{K}} \frac{\partial \log P}{\partial \log r}
\end{equation} 
with $P$ the gas pressure, given by:
\begin{equation}
P = \rho_{\rm{gas}} c_{s}^{2} = \frac{\Sigma_{\rm{gas}}}{\sqrt{2\pi} H_{\rm{gas}}} c_{s}^{2}
\end{equation}
with $\rho_{\rm{gas}}$ the midplane volume density of gas. 

\noindent We compare the coagulation and radial drift components of our model to well-known literature results in Appendix~A.

\subsection{Location of the snowline and the pebble ice fraction}\label{sec:snow}
As in SO17, we define the location of the snowline $r_{\rm{snow}}$ as the radius interior to which the water vapour pressure $P_{\rm{vap}}$ drops below the saturated (equilibrium) pressure $P_{\rm{eq}}$. Outside $r_{\rm{snow}}$, the water vapour distribution always follows the saturated profile, which is given by the Clausius-Clapeyron equation:
\begin{equation}\label{eq:peq}
P_{\text{eq}} = P_{\rm{eq, 0}} e^{- T_{a} / T}
\end{equation}
where $T_{a}$ and $P_{\rm{eq, 0}}$ are constants depending on the species. For water, $T_{a}$ = 6062 K and $P_{\rm{eq, 0}} = 1.14 \times 10^{13} \: \rm{g} \: \: \rm{cm}^{-1} \: \: \rm{s}^{-2}$ \citep{1991Icar...90..319L}. From \eq{peq} we find the saturated water vapour surface density profile $\Sigma_{\text{eq}}$:
\begin{equation}
\Sigma_{\text{eq}} = P_{\text{eq}} \frac{\sqrt{2 \pi} H_{\rm{gas}} \mu_{\rm{H}_{2} \rm{O}}}{k_{B} T}
\end{equation}
where $\mu_{\rm{H}_{2} \rm{O}}$ is the molecular weight of water. However, in contrast to SO17, in this work we do not follow the water vapour distribution directly. Instead, we calculate the expected water vapour surface density profile $\Sigma_{Z,\rm{a}}$ from the water mass flux that is delivered by icy pebbles $\dot{M}_{\rm{H}_{2} \rm{O}, \rm{peb}}$:
\begin{equation}\label{eq:vapeq}
\Sigma_{Z,\rm{a}} = \frac{\dot{M}_{\rm{H}_{2} \rm{O}, \rm{peb}}}{3 \pi \nu}
\end{equation}
which is equal to the steady-state advection-only vapor surface density profile from SO17\footnote{There is a typo in Eq.~(34) of SO17 for the steady-state advection-only vapor surface density profile. That equation should be the same as the current \eq{vapeq}.}. Note that $\Sigma_{Z,\rm{a}}$ is not the physical water vapour surface density profile, but rather the steady-state surface density profile if all the water in the disk is in the form of vapour. Under our assumptions, the physical water vapour profile would be given by the minimum of $\Sigma_{Z,\rm{a}}$ and $\Sigma_{\rm{eq}}$.

The location of the snowline $r_{\rm{snow}}$ can now be calculated by setting $\Sigma_{Z,\rm{a}}$ equal to $\Sigma_{\text{eq}}$ (SO17):
\begin{equation}\label{eq:snowloc}
\Sigma_{Z,\rm{a}} (r_{\rm{snow}}) = \Sigma_{\rm{eq}} (r_{\rm{snow}})
\end{equation}
Therefore, the location of the snowline depends on the pebble mass flux, which changes in time\footnote{If we take the accretion rate $\dot{M}_{\rm{gas}}$ to be a decreasing function of time and the snowline is located in the viscosity-dominated temperature region, the decrease in accretion heating is an additional effect that pushes the snowline inward \citep{2007ApJ...654..606G,2011ApJ...738..141O,2012MNRAS.425L...6M}. This is discussed in \se{mdottime}.}. In the simulations of SO17, the water mass flux was fixed and therefore the snowline did not move. In the current work, the water mass flux follows from the simulation: we determine $\Sigma_{Z,\rm{a}}$ from \eq{vapeq} by measuring $\dot{M}_{\rm{H}_{2} \rm{O}, \rm{peb}}$ just outside the snowline (at the innermost position where pebbles are icy) during the simulation. At the start of the simulation, the initial vapour surface density profile $\Sigma_{Z,\rm{a}}$ is given by $f_{\rm{ice, out}} \Sigma_{\rm{solids}} (r, t=0)$.

We define the ice fraction of pebbles $f_{\rm{ice}}$ as follows:
\begin{equation}\label{eq:icefr}
f_{\rm{ice}} = f_{\rm{ice, out}} \frac{(\Sigma_{Z,\rm{a}}^{4} + \Sigma_{\rm{eq}}^{4})^{1/4} - \Sigma_{\rm{eq}}}{\Sigma_{Z,\rm{a}}}
\end{equation}
where we have smoothed the maximum of $\Sigma_{Z,\rm{a}}$ and $\Sigma_{\rm{eq}}$ in order to avoid a sharp transition in the ice fraction profile. Using \eq{icefr} is a convenient way to calculate $f_{\rm{ice}}$ without having to model it directly by following the water vapour distribution.

In \fg{icefrac} we show the initial $\Sigma_{\rm{eq}}$ and $\Sigma_{Z,\rm{a}}$ profiles and the corresponding ice fraction profile $f_{\rm{ice}}$ for our fiducial model (\tb{inputpar}).

\begin{figure}[t]
	\centering
		\includegraphics[width=0.49\textwidth]{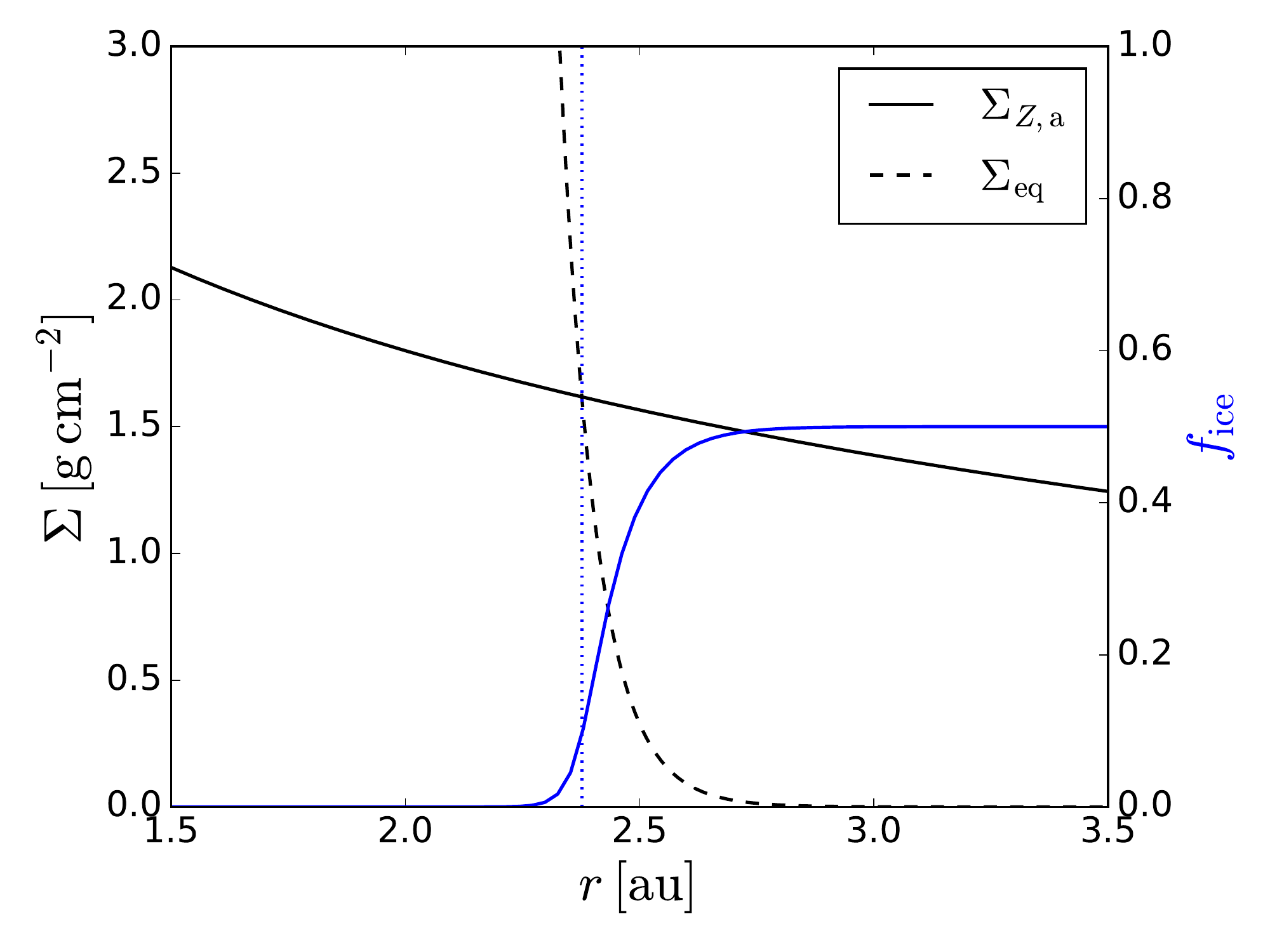}
\caption{Initial pebble ice fraction (blue solid line) as function of semi-major axis $r$ for the fiducial model. The saturated water vapour profile $\Sigma_{\rm{eq}}$ is plotted by the black dashed line, and $\Sigma_{Z,\rm{a}}$ (see text for more details) is given by the black solid line. The snowline is located where $\Sigma_{\rm{eq}}$ crosses $\Sigma_{Z,\rm{a}}$, in this case at $r_{\rm{snow}} \sim 2.4$ au (blue dotted line).\label{fig:icefrac}}
\end{figure}

\subsection{Diffusion of particles}\label{sec:raddif}
Our model does not directly account for radial diffusion of solid particles. However, the semi-analytic model that we employ for planetesimal formation outside the snowline (\se{SIsnowline}) does include turbulent radial diffusion of water vapour and of solid particles. Also, the radial velocity of solid particles (\eq{vdriftnoce}) includes the gas velocity component, meaning that if particles are small and well-coupled to the gas, they are moving towards the star at the gas accretion velocity. Therefore, for high values of $\alpha$, small particles in the inner disk can be in the so-called mixing regime \citep{2012A&A...539A.148B}, even if particle diffusion is not taken into account. Vertical diffusion of particles is accounted for by means of an $\alpha$- and $\tau$-dependent particle scale height (\eq{hsolids}).

\section{Planetesimal Formation by Streaming Instability}\label{sec:pltsml}
A promising planetesimal formation mechanism is streaming instability. Streaming instability occurs in the presence of radial drift, and leads to clumping of pebbles. These clumps can become dense enough to collapse under their own gravity and form planetesimals \citep{2005ApJ...620..459Y,2007Natur.448.1022J,2007ApJ...662..627J}. For streaming instability to be triggered, however, the solids-to-gas ratio needs to be locally enhanced above the typical, expected value of 1\% \citep{2009ApJ...704L..75J,BaiStone2010i,2015A&A...579A..43C,2017A&A...606A..80Y}. The condition for streaming instability that we adopt in this work is a midplane solids-to-gas mass density ratio ($\rho_{\rm{solids}} / \rho{\rm{gas}}$) exceeding unity \citep{DrazkowskaDullemond2014}, independent of the dimensionless stopping time of particles $\tau$. Several studies have found that the streaming instability threshold depends on $\tau$, with increasingly higher metallicities $Z \equiv \Sigma_{\rm{solids}} / \Sigma_{\rm{gas}}$ needed to trigger streaming instability for smaller values of $\tau$ ({\it e.g.}, \citet{2015A&A...579A..43C,2017A&A...606A..80Y}). However, those works considered laminar disks, whereas in our simulations the disks are intrinsically turbulent, so we cannot simply adopt their results. We find that the requirement of a midplane solids-to-gas ratio exceeding unity is always more constraining than the metallicity constraints found by \citet{2015A&A...579A..43C} and \citet{2017A&A...606A..80Y}. The behaviour of the streaming instability threshold on the metallicity with changing $\tau$ (increasing with decreasing $\tau$) is also captured by our choice of threshold, because the midplane solids-to-gas ratio decreases with decreasing $\tau$ for a given metallicity:
\begin{equation}
\frac{\rho_{\rm{solids}}}{\rho_{\rm{gas}}} = \frac{\Sigma_{\rm{solids}} H_{\rm{gas}}}{\Sigma_{\rm{gas}} H_{\rm{solids}}} = Z \sqrt{\frac{\tau + \alpha}{\alpha}}
\end{equation}
where in the last step we made use of \eq{hsolids}. The smallest particles that participate in planetesimal formation in our simulations have $\tau$$\sim$$10^{-3}$ (\se{directSI}), which according to \citet{2017A&A...606A..80Y} should indeed be able to trigger the streaming instability.

In this work, we are specifically (but not exclusively) interested in planetesimal formation around the water snowline. An enhanced solids-to-gas ratio can form interior to the snowline due to a traffic-jam effect caused by the variation of the fragmentation velocity across the snowline \citep{2011ApJ...728...20S,2016A&A...596L...3I}, as well as outside the snowline due to the effect of water vapour diffusion and condensation \citep{2004ApJ...614..490C,ArmitageEtal2016,SO2017,2017A&A...608A..92D}. In our model, we account for the effect of outward diffusion and re-condensation of water vapour by means of a recipe for planetesimal formation outside the snowline that we distill from SO17, since in this work we do not model the transport and condensation of water vapour directly.

Planetesimals are treated in a similar way as the dust particles: they are represented by super-particles containing planetesimals with the same physical characteristics. Planetesimal super-particles are characterised by a position, a total mass and a ice mass fraction. In contrast to pebble super-particles, the location of planetesimal super-particles is fixed; they are treated as sink particles. At the beginning of the simulation, a number $N_{\rm{pltsml}}$ planetesimal super-particles are initiated at a given location, with zero mass. If the solids-to-gas ratio at a certain radius exceeds unity (either directly or inferred from the `snowline recipe'), mass is transferred from the pebble super-particles to the nearest planetesimal super-particle.

\subsection{Direct streaming instability}\label{sec:direct}
Interior to the snowline, the solids-to-gas ratio can exceed unity `directly', due to a traffic-jam effect caused by the change in fragmentation velocity across the snowline. If this is the case, mass is transferred from pebbles to planetesimals at a rate $d M_{\rm{pl, SI, direct}} / d t$:
\begin{equation}\label{eq:directsi}
\frac{d M_{\rm{pl, SI, direct}}}{d t} = \frac{M_{\rm{peb, SI}}}{t_{\rm{settle}}} \epsilon_{\rm{SI}} = M_{\rm{peb, SI}} \epsilon_{\rm{SI}} \tau \Omega
\end{equation}
where $M_{\rm{peb, SI}}$ is the mass of the pebble super-particles that meet the requirement for SI, and $t_{\rm{settle}} = 1 / \tau \Omega$ is the timescale on which SI filaments form \citep{2017A&A...606A..80Y}, which equals the vertical settling timescale. The $\epsilon_{\rm{SI}}$ parameter is an efficiency parameter, that is considered a free parameter in some studies ({\it e.g.} \citet{DrazkowskaDullemond2014}). In our model, we take $\epsilon_{\rm{SI}} = 0.1$, and we come back to the relevance of this parameter in \se{directSI}.

\subsection{Outside (but near) the snowline}\label{sec:SIsnowline}
A local peak in the solids-to-gas ratio that exceeds unity is formed outside the snowline if the pebble mass flux $\dot{M}_{\rm{peb, outside}}$ outside the snowline (that is delivered from the outer disk) is larger than a certain critical value $\dot{M}_{\rm{peb, crit}}$, which depends on $\alpha$, $\dot{M}_{\rm{gas}}$, and on the dimensionless stopping time $\tau$ of particles outside the snowline (SO17). For a given disk model with constant $\alpha$ and $\dot{M}_{\rm{gas}}$, we employ the semi-analytic model presented in SO17 to calculate $\dot{M}_{\rm{peb, crit}}$ for different stopping times\footnote{The semi-analytic model of SO17 is not valid for dimensionless stopping times $\tau > 1$. We have updated it so that it can deal with $\tau > 1$ values as well. We present this update in Appendix~B.}. We assume the `many-seeds' scenario in this calculation, which leads to an ice fraction of planetesimals that is similar to that of pebbles in the outer disk: $f_{\rm{ice, pltsml}} = f_{\rm{ice, out}} = 0.5$ (\tb{inputpar}). This is because not only water vapour is transported across the snowline, but small silicate grains diffuse outward as well and stick to the inward-drifting pebbles (SO17).
The tabulated values of $\dot{M}_{\rm{peb, crit}}$ are then used to decide whether planetesimal formation takes place outside the snowline during the simulation. 

SO17 showed that the timescale $t_{\rm{peak, buildup}}$ on which the solids enhancement outside the snowline forms is related to the time it takes for water vapour to traverse the distance between the location where the solids-to-gas ratio peaks $r_{\rm{peak}}$, and the snowline location $r_{\rm{snow}}$. The width of the peak $W_{\rm{peak}} = r_{\rm{peak}} -  r_{\rm{snow}}$ depends on the disk parameters and is another outcome of the semi-analytic model.

The incoming pebble mass flux outside the snowline $\dot{M}_{\rm{peb, outside}}$ first increases as particles grow to pebble sizes. At some point it reaches a peak value after which it slowly decreases for the greater part of the lifetime of the disk (see also \citet{2014A&A...572A.107L}). 
As soon as $\dot{M}_{\rm{peb, outside}}$ exceeds the critical mass flux, we increase the planetesimal formation rate outside the snowline from zero to $d M_{\rm{pl, SI, snowline}} / d t$ on a peak formation timescale $t_{\rm{peak, buildup}} = W_{\rm{peak}} / |v_{\rm{gas}}|$. The value of $d M_{\rm{pl, SI, snowline}} / d t$ is given by:
\begin{equation}\label{eq:si}
\frac{d M_{\rm{pl, SI, snowline}}}{d t} = \max[\dot{M}_{\rm{peb}} - \dot{M}_{\rm{peb, crit}}, 0]
\end{equation}
where we assume that the peak in the solids-to-gas ratio is continuously sustained, and only the excess incoming pebble mass flux is transformed to planetesimals\footnote{One could imagine an `episodic' nature of streaming instability rather than the `continuous' approach we use here. In the episodic scenario, once streaming instability outside the snowline has been triggered the annulus of pebbles outside the snowline is emptied and need to be refilled by inward-drifting pebbles before streaming instability is triggered again, leading to episodic bursts of planetesimal formation. If the width of the peak $W_{\rm{peak}}$ is much larger than $\eta r$ (the distance a drifting pebble traverses during one vertical settling timescale) and all pebbles in the peak participate in planetesimal formation, the episodic description may be better than the continuous one. However, we checked that the timescales involved in emptying and refilling the snowline region with pebbles are much shorter than the planetesimal formation phase, which justifies the continuous approach.}. The mass of the planetesimal super-particle closest to the peak location $r_{\rm{peak}}$ (which is given by the semi-analytic model) is increased at a rate $d M_{\rm{pl, SI, snowline}} / d t$. 

We distribute the pebble mass loss rate over all pebble super-particles using a Gaussian kernel centred on the peak location $r_{\rm{peak}}$. The mass loss rate of pebble super-particle $i$ is then given by:
\begin{equation}\label{eq:pebSI}
\frac{d M_{\rm{peb, SI, snowline, i}}}{d t} = f(r_{i}) \frac{d M_{\rm{pl, SI, snowline}}}{d t}
\end{equation}
where the fraction $f(r_{i}) < 1$ depends on the distance between $r_{i}$ and the peak location\footnote{To be specific, \eq{pebSI} reads:
\begin{multline}\label{eq:pebmassloss}
\frac{d M_{\rm{peb, SI, snowline}}}{d t} (r_{i}) = \\ [(1 - f_{\rm{ice, out}}) + f_{\rm{ice}} (r_{i})] \frac{W_{\rm{Gauss}} (r_{i}, r_{\rm{peak}})}{\sum_{j = 0}^{N_{\rm{peb}}} W_{\rm{Gauss}} (r_{j}, r_{\rm{peak}})} \frac{d M_{\rm{pl, SI, snowline}}}{d t}
\end{multline}
where $N_{\rm{peb}}$ is the number of pebble super-particles and $W_{\rm{Gauss}}$ is given by:
\begin{equation}\label{eq:gauss}
W_{\rm{Gauss}} (r, r_{\rm{c}}) = \frac{1}{0.1 r_{\rm{c}} \sqrt{2 \pi}} \exp\left[-\frac{1}{2} \left(\frac{r - r_{\rm{c}}}{0.1 r_{\rm{c}}}\right)^{2}\right]
\end{equation}
The term $[(1 - f_{\rm{ice, out}}) + f_{\rm{ice}}]$ in \eq{pebmassloss} takes into account the fact that in our model the location of the solids enhancement outside the snowline $r_{\rm{peak}}$ can be at a distance from the star where $f_{\rm{ice}} < f_{\rm{out}}$, again because we do not follow the transport of water vapour explicitly in this work.}.

Throughout the simulation, we make sure to measure $\dot{M}_{\rm{peb, outside}}$ outside the region where pebbles are converted to planetesimals.

The total mass that is lost from the pebbles (disregarding the term $[(1 - f_{\rm{ice, out}}) + f_{\rm{ice}}]$; see footnote 6) is of course the exact negative of \eq{si}, such that when streaming instability outside the snowline is going on (and the solids enhancement outside the snowline has saturated), the pebble mass flux that reaches the actual snowline is $\dot{M}_{\rm{peb, crit}}$. A consequence of this is that the location of the snowline is fixed during planetesimal formation outside the snowline (\se{snow}).

Once the pebble flux has dropped below the critical value $\dot{M}_{\rm{peb, crit}}$, a peak in the midplane solids-to-gas density ratio that exceeds unity cannot be sustained outside the snowline any longer, and the pebble mass flux that reaches the snowline is again the same as the pebble mass flux outside the snowline, because just outside the snowline no pebbles are converted to planetesimals anymore.

\begin{figure*}[t]
	\centering
		\includegraphics[width=\textwidth]{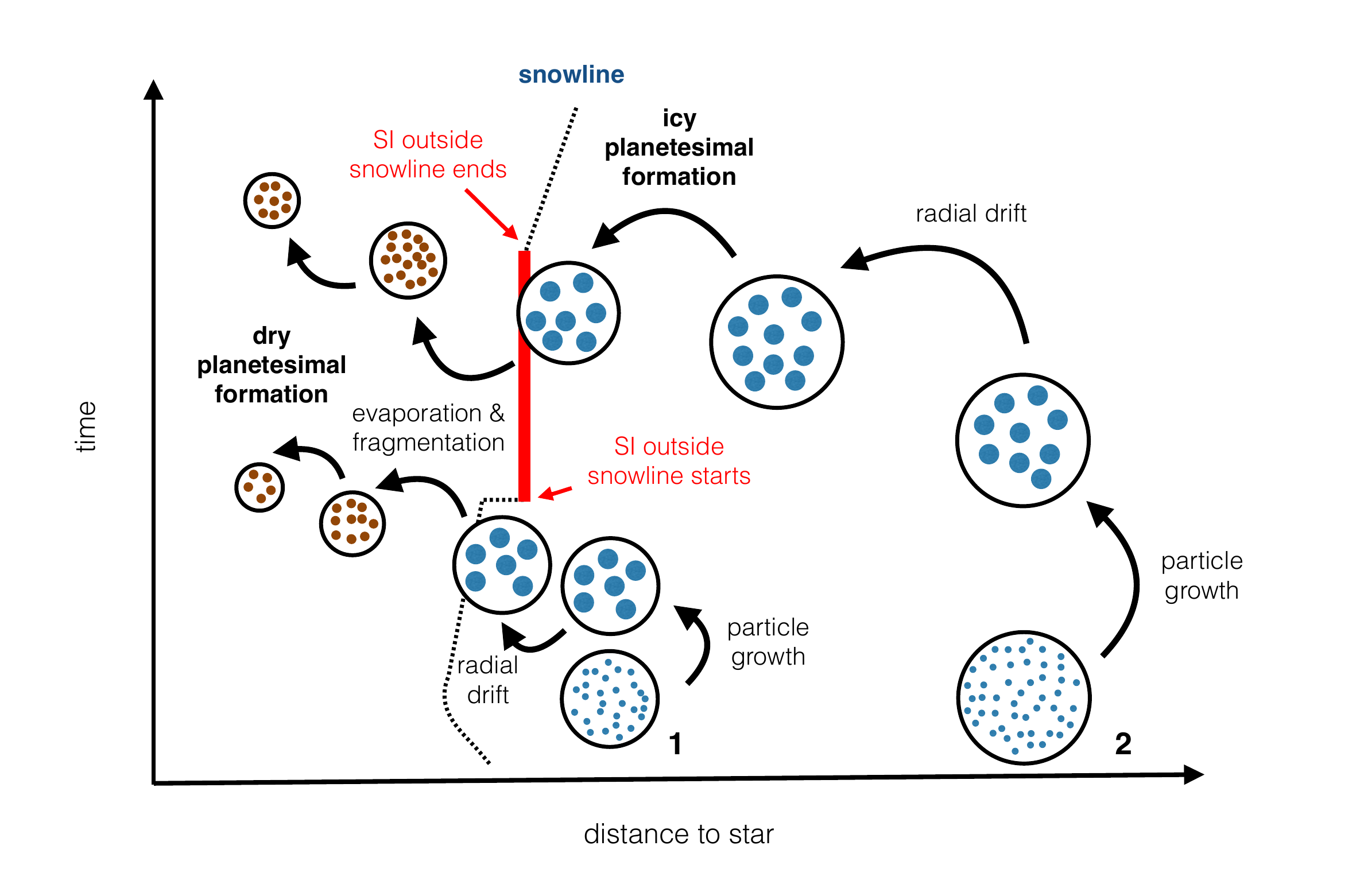}
\caption{Cartoon of two possible trajectories through space and time of pebble super-particles ({\bf 1} and {\bf 2}). Both super-particles start out in the outer disk with a small typical particle mass. The physical particles represented by the super-particles grow until they reach a size where radial drift becomes faster than growth, and they start to drift inward. Planetesimal formation occurs outside the snowline as well as in the dry inner disk. See text for a more elaborate description.\label{fig:cartoon}}
\end{figure*}

\section{Model Summary}\label{sec:summary}
To summarise the various effects that we take into account in our model, \fg{cartoon} shows a schematic of two possible `lifelines' (trajectories through space and time) of pebble super-particles that start in the icy outer disk and eventually convert part of their mass to planetesimals. 
The first super-particle (denoted by a `1' in the cartoon) begins its journey just outside the snowline, representing small icy particles of 0.1 micron (\tb{inputpar}). Once these particles have grown to pebble sizes, the super-particle starts to drift inward (\eq{vdriftnoce}). At the time when it reaches the water snowline, the conditions for streaming instability outside the snowline are not yet met (the pebble mass flux does not exceed the critical mass flux yet; \eq{si}). Therefore, the super-particle does not lose any mass to planetesimals outside the snowline. As the super-particle crosses the snowline, its $\rm{H}_{2}\rm{O}$ ice evaporates (\eq{icefr}), and the physical particles it represents fragment down to smaller sizes during the snowline crossing because of the lower fragmentation threshold for silicate particles (\eq{fragvel}). The inner disk is fragmentation-limited, which leads to a steeper radial dependence of the solids surface density than of the gas surface density \citep{2012A&A...539A.148B}, and therefore to an increasing solids-to-gas ratio with decreasing distance to the star. In this cartoon the solids-to-gas ratio reaches unity close to the star, resulting in dry planetesimal formation (\eq{directsi}).

The second super-particle (indicated by a `2' in the cartoon) starts out further away from the star, also representing 0.1 micron-sized icy particles. These particles grow to pebble-sizes at a slower pace than the particles of super-particle 1, because the growth timescale (\eq{massgrowthrate}) is longer for larger distances from the star (because the solids density goes down with increasing semi-major axis). At the time when the super-particle reaches the snowline by radial drift, the conditions for streaming instability outside the snowline are met, and part of its mass is converted to planetesimals before it crosses the snowline (\eq{si}). There, the water ice content of the super-particle evaporates and the physical particles that it represents fragment to smaller sizes.

During the evolution of the solids, the position of the snowline changes. At first, it is pushed closer to the star as the pebble mass flux increases, after which it slowly recedes as the pebble mass flux decreases (\eq{snowloc}). When streaming instability outside the snowline starts (indicated with `SI outside snowline starts'), the pebble mass flux reaching the snowline is reduced because part of it is converted to planetesimals outside the snowline. When the pebble mass flux outside the snowline falls below the critical pebble mass flux for streaming instability, planetesimal formation outside the snowline stalls, and the snowline resumes to recede as the pebble mass flux delivered from the outer disk continues to decrease.

\section{Fiducial Model Results}\label{sec:fid}
In this section we present results for our fiducial model, where we consider a protoplanetary disk around a solar-mass star. The parameter values for the fiducial model can be found in Table~1. We constrain the total disk mass $M_{\rm{disk}}$ to $0.04 M_{\odot}$. Together with the fixed values for $\alpha$ and $\dot{M}_{\rm{gas}}$, this constrains the outer radius of the disk $r_{\rm{out}}$, which for the fiducial model is 55 au.

\Fg{mparticle} shows the particle mass $m_{p}$ as function of semi-major axis $r$ at different time points. The red line indicates the fragmentation limit on the mass (see \se{frag}) at the last time point plotted ($5 \times 10^{5}$ years). After a few hundred years, particles interior to the snowline have grown to their fragmentation limit. In the outer part of the disk ($\gtrsim$10 au) the fragmentation limit is never reached; there the drift timescale is shorter than the growth timescale. The boundary between the fragmentation-limited and the drift-limited regions gets closer to the star in time, as the solids surface density goes down.

In \fg{sigma_pltsmls_2} we plot the solids surface density profiles at different points in time. After a few thousand years, a `traffic jam' materialises interior to the snowline. This pile-up then spreads throughout the inner disk because an icy pebble mass flux from the outer disk keeps delivering material to the snowline region. At some point, as the pile-up is extending through the inner disk, the pebble mass flux that is delivered interior to the snowline starts to decrease, and the solids surface density interior to the snowline goes down as well. The decrease of the pebble mass flux is exacerbated by the streaming instability, because of which pebbles are converted to planetesimals outside the snowline. This is made clear in \fg{massflux_pltsmls}. The surface density profiles in \fg{sigma_pltsmls_2} also show the transition from the fragmentation-dominated phase to the drift-dominated phase outside the snowline. For $t < 10^{5} \: \rm{yr}$, the region just outside the snowline is still fragmentation-dominated (solids surface density $\propto r^{-3/2}$; \citet{2012A&A...539A.148B}). At $t = 2 \times 10^{5} \: \rm{yr}$, the region outside the snowline has become drift-dominated, which is shown by an $r^{-1}$ surface density power law.

\Fg{massflux_pltsmls} shows the pebble mass flux that arrives at the snowline $r_{\rm{snow}}$ as a function of time (solid black line), as well as the pebble mass flux that arrives outside the planetesimal annulus (dashed black line) and the streaming instability threshold on the pebble mass flux (dotted line). The total mass in planetesimals is plotted in blue. The pebble mass flux first increases as particles grow (\fg{mparticle}). The critical pebble mass flux needed for streaming instability varies with the stopping time of particles outside the snowline, and reaches a constant value when particles outside the snowline have reached their fragmentation limit, at about $10^{3}$ years. When the pebble mass flux exceeds the critical pebble mass flux, streaming instability operates and pebbles are being converted to planetesimals at a rate given by \eq{si}. The pebble mass flux reaching the snowline is decreased to the critical pebble mass flux on the peak formation timescale (\se{SIsnowline}). The pebble mass flux that is still reaching the snowline after the enhancement has saturated, during the planetesimal formation phase, is equal to the critical pebble mass flux $\dot{M}_{\rm{peb, crit}}$ that is needed to sustain the high solids-to-gas ratio outside the snowline (\se{SIsnowline}). When the pebble mass flux that is delivered to the streaming instability region (outside $r_{\rm{snow}}$; dashed line) drops below $\dot{M}_{\rm{peb, crit}}$, streaming instability shuts off and the pebble mass flux reaching $r_{\rm{snow}}$ is again equal to the pebble mass flux reaching the planetesimals annulus outside $r_{\rm{snow}}$. This happens at around $3.2 \times 10^{4}$ years. The total yield in planetesimals is about ten Earth masses.

\begin{table}
\caption{Input parameters, their symbols, and their fiducial values.}
\label{tab:inputpar}
\centering
\begin{tabular}{l l l}
\hline
Stellar mass & $M_{\star}$ &$1 \: M_{\odot}$\\
Gas accretion rate & $\dot{M}_{\rm{gas}}$ &$5 \times 10^{-9} \: M_{\odot} \: \rm{yr}^{-1}$\\
Total disk mass & $M_{\rm{disk}}$ & $0.04 \: M_{\odot}$\\
Disk outer radius & $r_{\rm{out}}$ & 55 \rm{au}\\
Metallicity & $Z$ &0.01\\ 
Ice fraction outer disk & $f_{\rm{ice, out}}$ & 0.5\\
Initial particle size & $s_{\rm{p}}$ & $0.1 \: \mu m$\\
Turbulence strength & $\alpha$ & $10^{-3}$\\
Fragm. threshold, icy & $v_{\rm{frag, icy}}$ &$10 \: \rm{m} \: \rm{s}^{-1}$\\
Fragm. threshold, silicate & $v_{\rm{frag, dry}}$ &$3 \: \rm{m} \: \rm{s}^{-1}$\\
Viscous temp. at 1 au & $T_{\rm{visc, 1 au}}$ &$350 \: \rm{K}$\\
Irradiation temp. at 1 au & $T_{\rm{irr, 1 au}}$ &$177 \: \rm{K}$\\
\hline
\end{tabular}
\end{table}

\begin{figure}[t]
	\centering
		\includegraphics[width=0.49\textwidth]{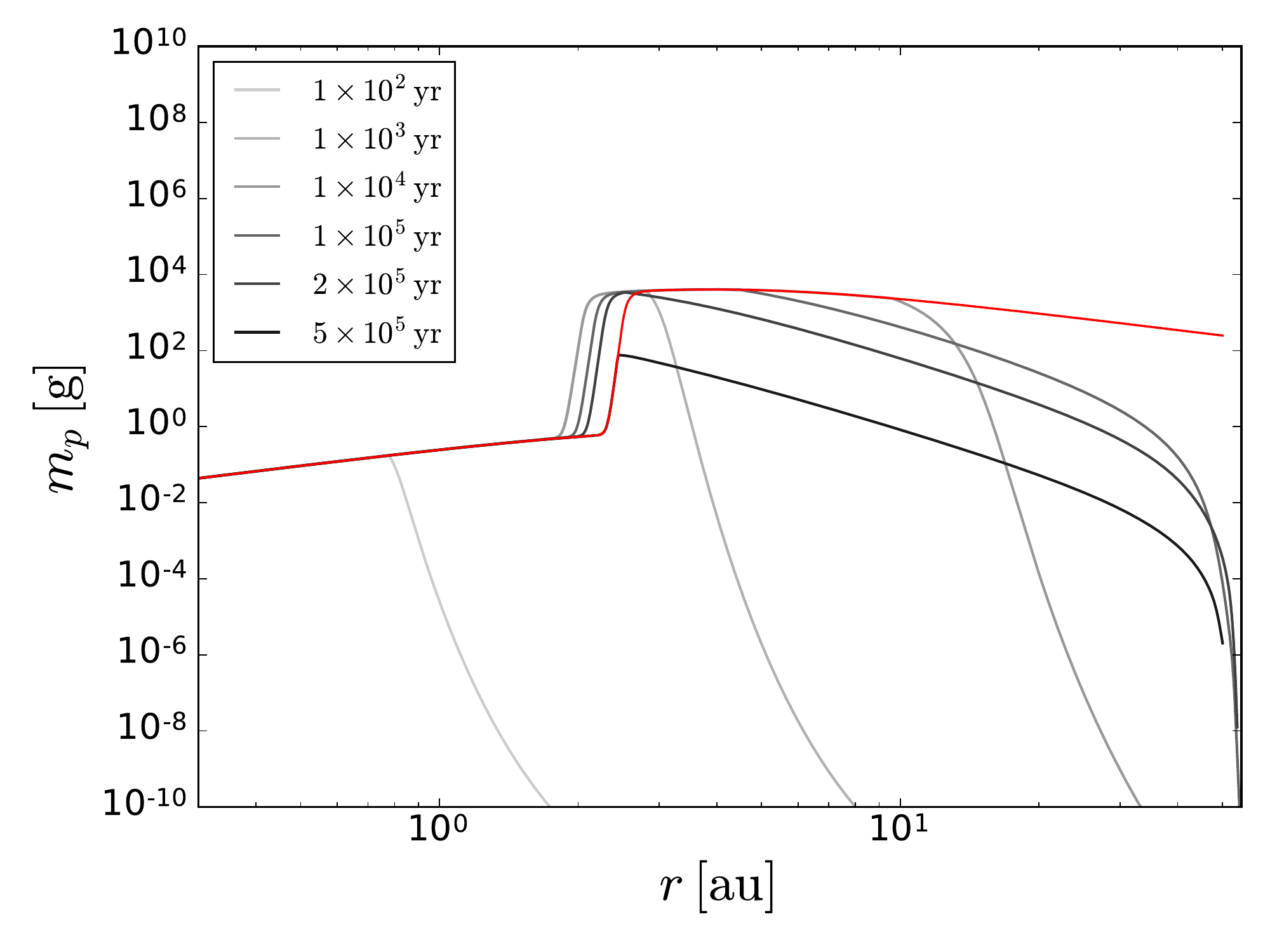}
\caption{Particle mass $m_{p}$ as function of semi-major axis $r$ at different points in time for the fiducial model (1a). The red line corresponds to the fragmentation limit on the mass (see \fg{mfrag}) for the last time point plotted ($t = 5 \times 10^{5} \: \rm{yr}$).\label{fig:mparticle}}
\end{figure}

\begin{figure}[t]
	\centering
		\includegraphics[width=0.49\textwidth]{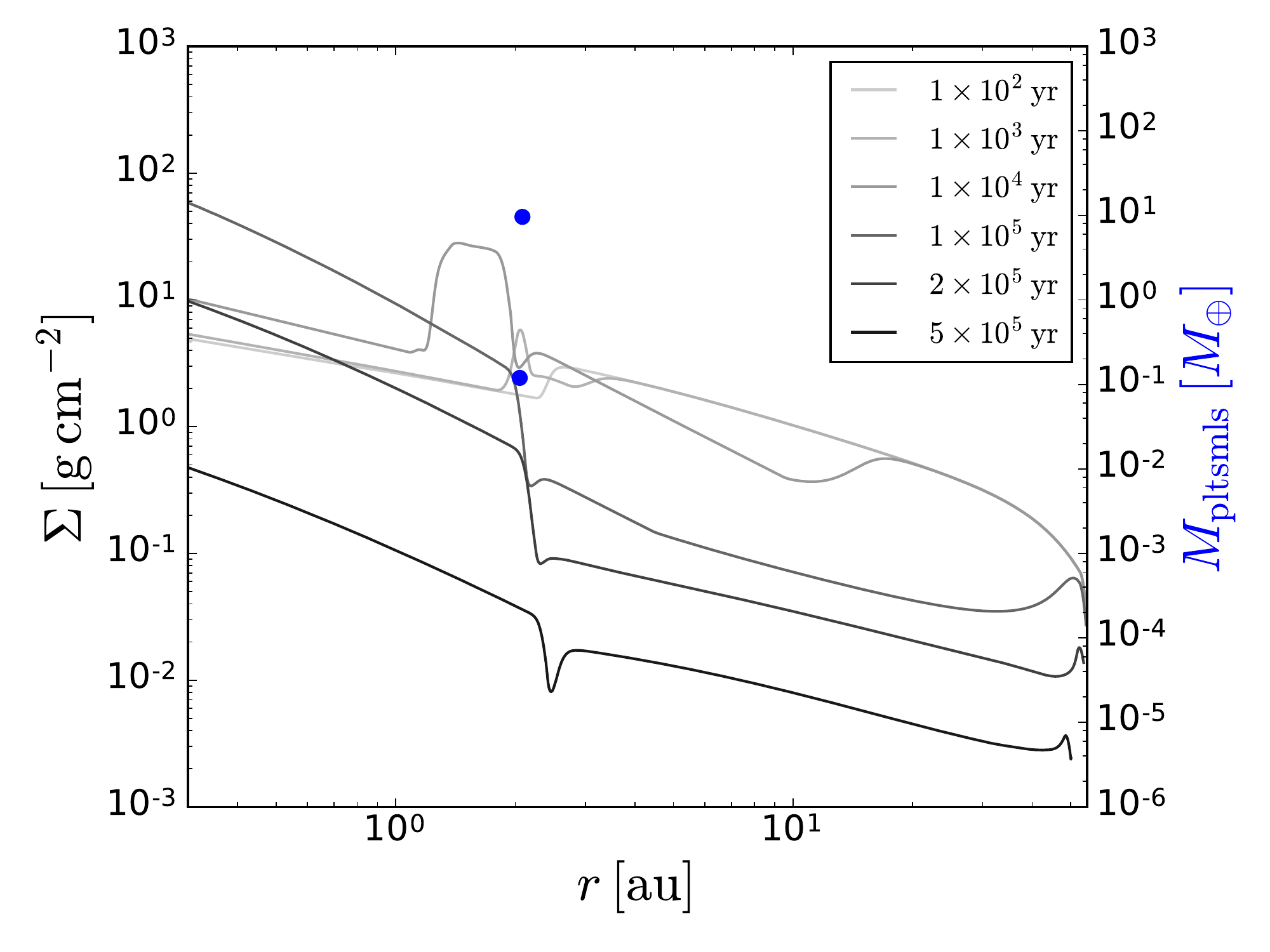}
\caption{Solids surface density profiles $\Sigma$ at different points in time for the fiducial model (1a). The blue dots correspond to the mass of planetesimal super-particles $\dot{M}_{\rm{pltsmls}}$ in Earth masses at the last time point plotted ($t = 5 \times 10^{5} \: \rm{yr}$) (this is also the final total mass in planetesimals, because the planetesimal formation phase ends at $\sim$$3.2 \times 10^{4} \: \rm{yr}$).\label{fig:sigma_pltsmls_2}}
\end{figure}

\begin{figure}[t]
	\centering
		\includegraphics[width=0.49\textwidth]{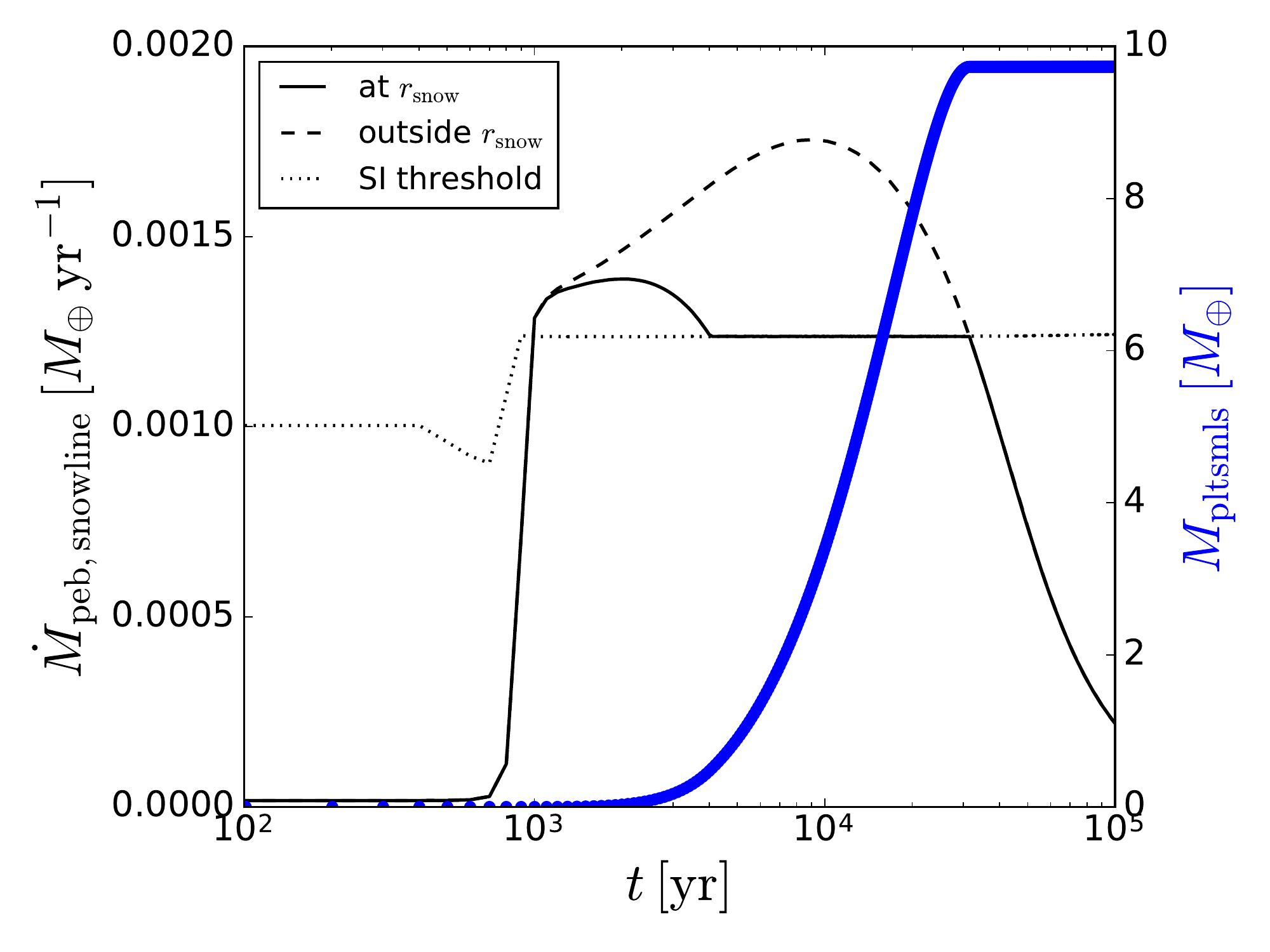}
\caption{Pebble mass flux $\dot{M}_{\rm{peb}}$ that arrives at the snowline $r_{\rm{snow}}$ in Earth masses per year, as a function of time $t$ (solid black line), for the fiducial model (1a). The dashed black line corresponds to the pebble mass flux $\dot{M}_{\rm{peb}}$ that arrives at the planetesimal annulus outside $r_{\rm{snow}}$. During the planetesimal formation phase, the pebble mass flux at the snowline is constant and equal to the pebble mass flux delivered by the `peak' outside the snowline $\dot{M}_{\rm{peb, crit}}$ (dotted line): the excess flux is converted to planetesimals. The total mass in planetesimals $M_{\rm{pltsmls}}$ is plotted in blue.\label{fig:massflux_pltsmls}}
\end{figure}

\section{Parameter Study Results}\label{sec:parresults}
We investigate planetesimal formation in disks around stars of different masses. We scale the gas accretion rate $\dot{M}_{\rm{gas}}$ with stellar mass $M_{\star}$ as \hbox{$\dot{M}_{\rm{gas}} \propto M_{\star}^{2.2}$}, consistent with observations \citep{2017A&A...604A.127M,2017ApJ...847...31M}. The viscous temperature $T_{\rm{visc}}$ scales with $(\dot{M}_{\rm{gas}} M_{\star})^{1/4}$, where $M_{\star}$ is the stellar mass \citep{2002apa..book.....F}. It then follows that $T_{\rm{visc}} \propto M_{\star}^{4/5}$.
We use the following mass-luminosity relation: \hbox{$L_{\star} \propto M_{\star}^{4}$}, with which the irradiation temperature $T_{\rm{irr}}$ scales linearly with $M_{\star}$.

\noindent
In our steady-state viscous gas disk model, the parameters $\dot{M}_{\rm{gas}}$ and $\alpha$ determine the gas surface density profile $\Sigma_{\rm{gas}}$ (\eq{svis}). One then needs to either fix the total disk mass $M_{\rm{disk}}$ or the outer disk radius $r_{\rm{out}}$ to find the other. In constructing disk models around different stellar masses for our parameter study, we consider two scenarios. In the first one, we constrain $M_{\rm{disk}}$ to $4\% \: M_{\star}$, and the value of $r_{\rm{out}}$ follows. In the second one, we take $r_{\rm{out}} = 150 \: \rm{au}$, from which the value of $M_{\rm{disk}}$ follows. 
We omit disks that would have $M_{\rm{disk}} > 10 \% \: M_{\star}$ or $M_{\rm{disk}} < 1 \% \: M_{\star}$, as well as disks for which $r_{\rm{out}} > 500 \: \rm{au}$ or  $r_{\rm{out}} < 30 \: \rm{au}$. 

In \tb{parstudy} we list the parameter values for the different disk models, as well as the results (total masses in icy and dry planetesimals, and the time period during which streaming instability outside the snowline is operational (if it is operational at all)). We run models with stellar masses $M_{\star}$ of $0.1 \: M_{\odot}$, $0.5 \: M_{\odot}$, $1 \: M_{\odot}$ and $2 \: M_{\odot}$ (model IDs starting with 2, 3, 1, and 4, respectively).
We also included models where the gas accretion rate $\dot{M}_{\rm{gas}}$ varies with time; where we increased the metallicity $Z$; where we include pebble accretion; and where we increase the fragmentation velocity for icy particles $v_{\rm{frag, ice}}$. These models are discussed in \se{mdottime}, \se{directSI}, \se{pa}, and \se{driftlim}, respectively.

\subsection{Varying stellar mass}
Let us first compare the results of models 1a, 2b, 3a, and 4a to take a look at the stellar mass dependence for a fixed value of $\alpha = 10^{-3}$. We find that models 1a, 2b, and 3a (corresponding to 1, 0.1, and 0.5 solar-mass stars, respectively) convert similar percentages of their initial solids reservoir to planetesimals ($\sim$$10\%$). Model 4a ($M_{\star} = 2 M_{\odot}$) does not form any planetesimals. The critical pebbles-to-gas mass flux ratio $\dot{M}_{\rm{peb, crit}} / \dot{M}_{\rm{gas}}$ that is required for streaming instability depends only weakly on the values of $\alpha$ and $\dot{M}_{\rm{gas}}$. We find however that the ratio of the actual pebbles-to-gas mass flux ratio $\mathcal{F}_{s/g}$ just outside the snowline does depend on the stellar mass, through the snowline location. We can write $\mathcal{F}_{{\rm s/g}}$ as:
\begin{equation}\label{eq:fsgsnow}
\mathcal{F}_{{\rm s/g}} \equiv \frac{\dot{M}_{\rm{peb, snow, out}}}{\dot{M}_{\rm{gas}}} \propto \frac{\Sigma_{\rm{peb, snow, out}} v_{p} r_{\rm{snow, out}}}{\dot{M}_{\rm{gas}}}
\end{equation}
where here $r_{\rm{snow, out}}$ is the location just outside the snowline, $\Sigma_{\rm{peb, snow, out}}$ is the pebble surface density at $r_{\rm{snow, out}}$, and $v_{p}$ is the drift velocity of pebbles at $r_{\rm{snow, out}}$. Making use of the relation $\Sigma_{\rm{gas}} \propto \dot{M}_{\rm{gas}} \alpha^{-1}$, and neglecting effects such as the time-evolution of the solids disk and the peak formation timescale, to zeroth order we can approximate \eq{fsgsnow} as:
\begin{equation}\label{eq:snowlinedependence}
\mathcal{F}_{{\rm s/g}} \propto \alpha^{-1} r_{\rm{snow}}^{1 + p} \tau
\end{equation}
where $\tau$ is the dimensionless stopping time of pebbles just outside the snowline (where particles are icy) and $p$ is the power-law index of the pebbles surface density. In case the region outside the snowline is fragmentation-limited, as is the case in our simulations (except for model 1l; see \se{driftlim}); $p = -3/2$. In the drift-limited case $p = -1$. Assuming that the temperature at the snowline is independent of stellar mass, the fragmentation limit on $\tau$ is proportional to $\alpha^{-1}$ \citep{2012A&A...539A.148B}. The drift limit on $\tau$ is proportional to $(V_{K} / c_{s})^{2}$ with $V_{K}$ the Kepler velocity and $c_{s}$ the sound speed \citep{2012A&A...539A.148B}, which boils down to $\tau \propto r_{\rm{snow}}^{-1}$ if we again assume a snowline temperature independent of stellar mass. We then find $\mathcal{F}_{{\rm s/g}} \propto \alpha^{-1} r_{\rm{snow}}^{-1}$ if the region outside the snowline is drift-limited and $\mathcal{F}_{{\rm s/g}} \propto \alpha^{-2} r_{\rm{snow}}^{-1/2}$ for the fragmentation-limited case. In both cases, the pebbles-to-gas mass flux ratio that is important for planetesimal formation becomes smaller for more massive stars that have their snowline further away.
Because the snowline is located further away in model 4a than in models 1a, 2b, and 3c, the critical pebble mass flux is (just) not reached in this disk and no planetesimals are formed.

The negative relation between $\mathcal{F}_{{\rm s/g, snow}}$ and $\alpha$ (partly due to the set-up of the $\alpha$-viscosity gas disks ($\Sigma_{\rm{gas}} \propto \dot{M}_{\rm{gas}} \alpha^{-1}$) as was already found in SO17), is clearly seen in \tb{parstudy}. For a given stellar mass, the lower $\alpha$, the higher the total pebbles-to-planetesimal conversion efficiency. Model 4a has the lowest $\alpha$ value among all runs with $M_{\star} = 2 M_{\odot}$, and comes closest to forming planetesimals: the pebble mass fluxes in the other disks in batch 4, which have higher $\alpha$ values, are even lower.

We have to keep in mind though that the zeroth-order dependencies on the stellar mass discussed above can be overshadowed by the effects of the gas disk. We have already discussed the effect of the value of $\alpha$ (which by no means needs to be constant throughout the disk as assumed in this work); but the transition from the Epstein to the Stokes drag regime; the compactness of the disk; deviations from the viscously-relaxed gas profile; the time when the fragmentation-limited region outside the snowline becomes drift-limited, etcetera, could play important roles as well. For example, one might notice that model 3a has a higher pebbles-to-planetesimals conversion efficiency than model 2b, which is not expected from the snowline-dependency argument alone. However, disk 3a is much more compact than disk 2b. This means that disk 2b has a lot of solids material in the outer disk that is not used for streaming instability, because at the time when pebbles that originate in the outer disk reach the snowline, the planetesimal formation phase has already ended. This reduces the conversion efficiency in model 2b.
The compactness of a disk can also be a limiting factor to the total amount of planetesimals that can form. Model disk 3d is the same as model disk 3a except that it is larger. Model 3d forms more planetesimals than model 3a because a high enough pebble mass flux is sustained for a longer period of time (but again, the total conversion efficiency is lower).
Therefore, even though the pebbles-to-gas mass flux ratio outside the snowline $\mathcal{F}_{{\rm s/g}}$ depends on the snowline location, we cannot directly translate this dependency to a simple scaling law for the pebbles-to-planetesimals conversion efficiency. Such secondary effects are precisely the reason why fast and versatile planetesimal formation models such as the one presented here are necessary.

\subsection{Time-dependent accretion rate}\label{sec:mdottime}
\begin{figure}[t]
	\centering
		\includegraphics[width=0.49\textwidth]{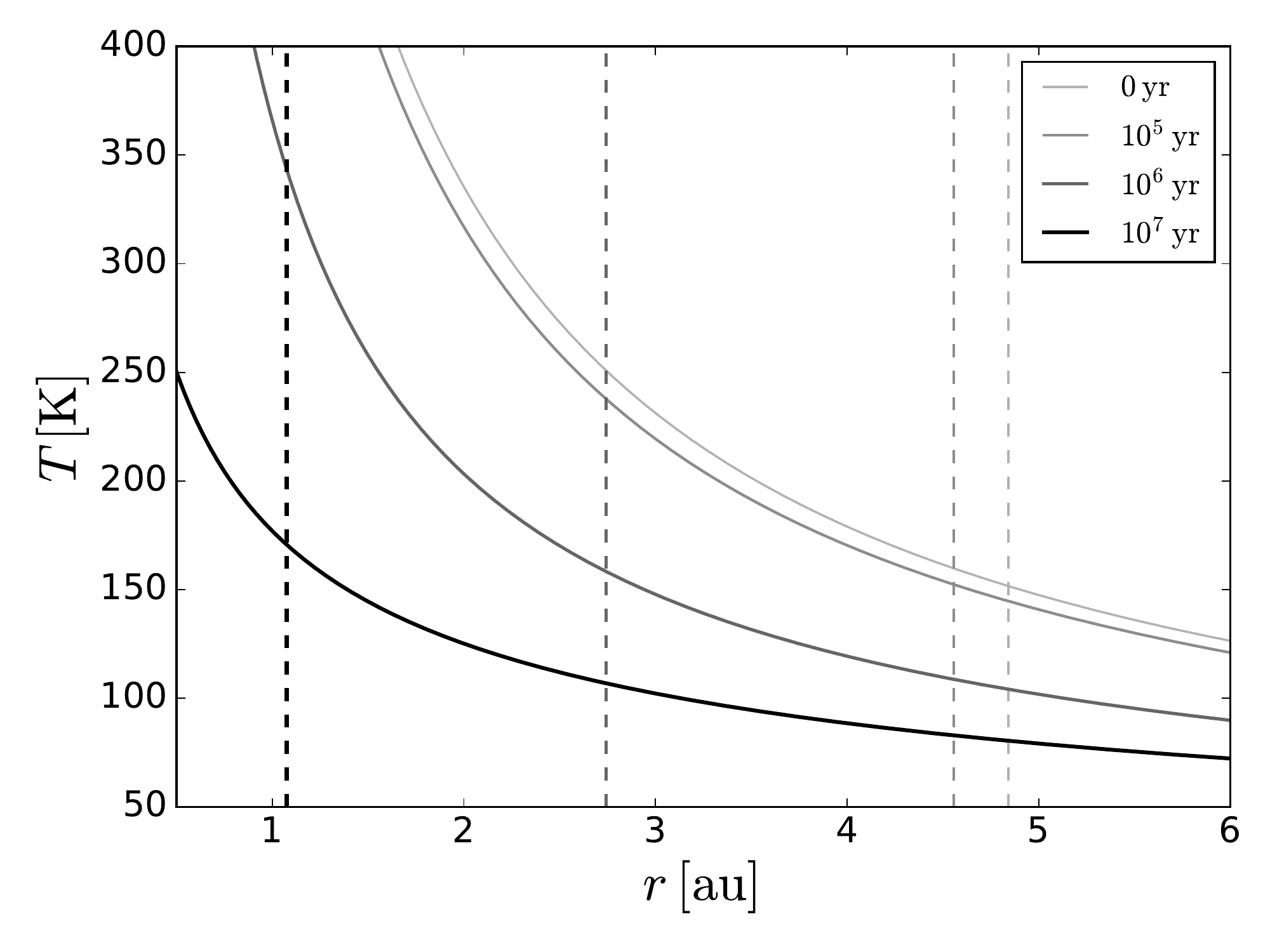}
\caption{Temperature $T$ as function of semi-major axis $r$ at different points in time, for model 1g (\tb{parstudy}), where the accretion rate starts out at \hbox{$\dot{M}_{\rm{gas}} = 5 \times 10^{-8} \: M_{\odot} \: \rm{yr}^{-1}$} and decreases on the viscous evolution timescale (see text for more details). Time is denoted in years, and the vertical dashed lines indicate the corresponding snowline locations for a constant, arbitrary ice flux $\dot{M}_{\rm{H}_{2}\rm{O}} = 3 \times 10^{-4} M_{\oplus} \: \rm{yr}^{-1}$. \label{fig:icelineloc}}
\end{figure}

\begin{figure}[t]
	\centering
		\includegraphics[width=0.49\textwidth]{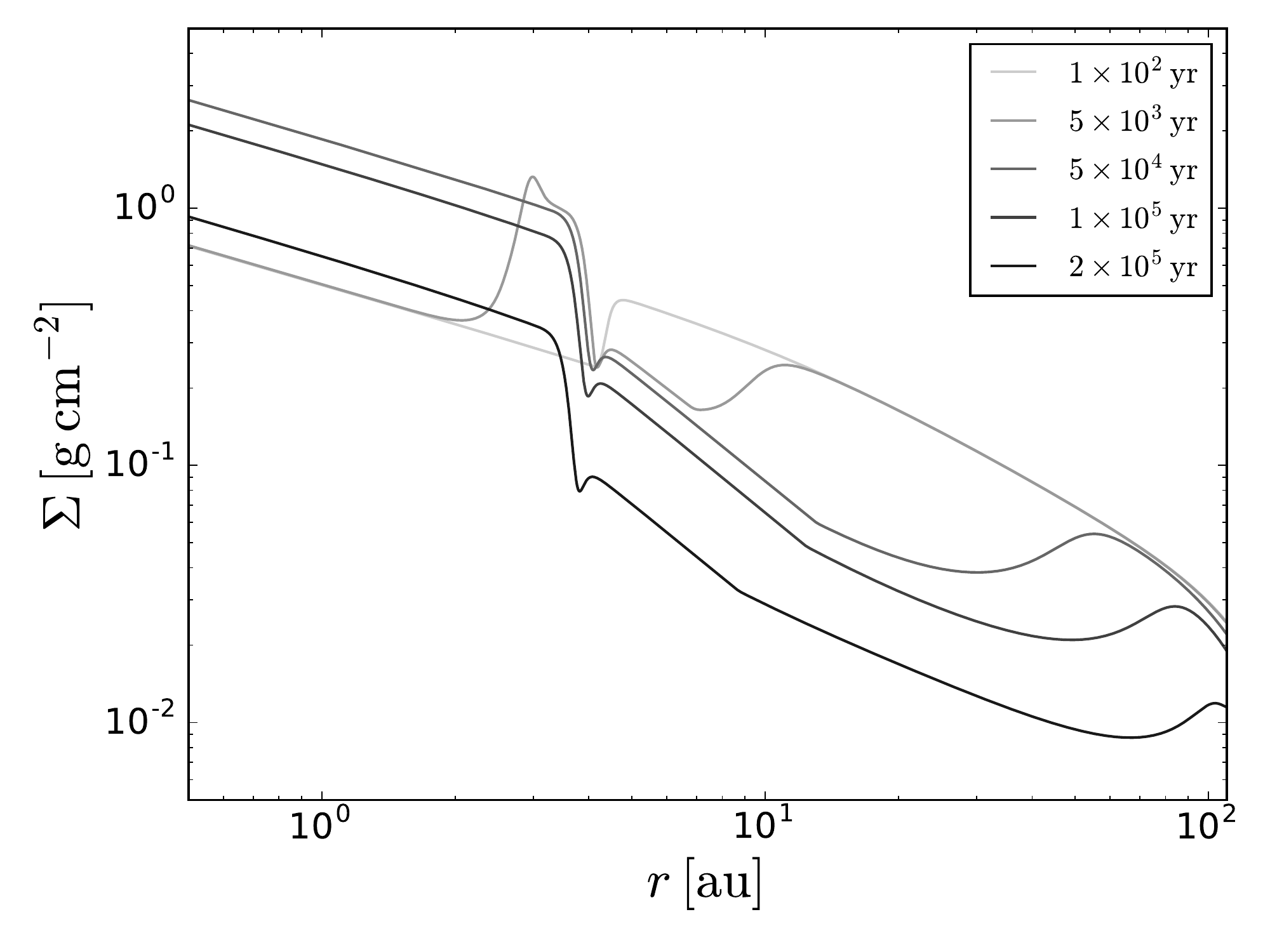}
\caption{Solids surface density profiles $\Sigma$ at different points in time, for model 1g (\tb{parstudy}). In this model the gas accretion rate is time-dependent and $\alpha = 3 \times 10^{-2}$. No planetesimals have formed.\label{fig:viscsigmasolids}}
\end{figure}

Up to now, we have kept $\dot{M}_{\rm{gas}}$ constant in our model runs. In reality, the gas accretion rate $\dot{M}_{\rm{gas}}$ decreases on a viscous evolution timescale $t_{\rm{visc}} = r_{1}^{2} / \nu_{1}$, where $r_{1}$ is the characteristic radius of the disk and $\nu_{1}$ is the viscosity at $r_1$ \citep{1974MNRAS.168..603L,1998ApJ...495..385H}. The evolution of the solids content of the disk proceeds faster than $t_{\rm{visc}}$, and therefore a constant $\Sigma_{\rm{gas}}$ profile is justified. However, the location of the water snowline, which is an important quantity in our model, is highly sensitive to the temperature structure in the disk. The temperature in the inner disk, which is dominated by viscous heating, depends on $\dot{M}_{\rm{gas}}$. Therefore a decrease in $\dot{M}_{\rm{gas}}$ ---however small--- could lead to an inward movement of the water snowline during the planetesimal formation phase \citep{2007ApJ...654..606G,2011ApJ...738..141O,2012MNRAS.425L...6M}. 
The viscous evolution timescale is shorter for larger $\alpha$. Therefore, to investigate how viscous evolution affects our results, we run a model with a high $\alpha$-value of $3 \times 10^{-2}$, where the viscous temperature $T_{\rm{visc}}$ goes down with decreasing $\dot{M}_{\rm{gas}}$ (\se{temp}). The gas surface density profile $\Sigma_{\rm{gas}}$ is adjusted accordingly, but we do not account for viscous spreading of the disk (we keep $r_{\rm{out}}$ fixed). This is not a problem, because it would not change the situation around the snowline and the inner disk. Because of the high $\alpha$-value, we also increase the fragmentation velocity for icy particles $v_{\rm{frag, icy}}$ from $10 \: \rm{m} \: \rm{s}^{-1}$ to $60 \: \rm{m} \: \rm{s}^{-1}$ (see also \se{driftlim}), to get non-negligible drift velocities. The time-dependent accretion rate $\dot{M}_{\rm{gas}} (t)$ is given by:
\begin{equation}
\dot{M}_{\rm{gas}} (t)= 5 \times 10^{-8} \exp{[-t / t_{\rm{visc}}]} \: M_{\odot} \: \rm{yr}^{-1}
\end{equation}
where we take for the characteristic radius of the disk $r_{1} = 50 \: \rm{au}$.

In \fg{icelineloc} we plot the resulting temperature profile at different points in time, as well as the corresponding snowline locations for a given, arbitrary ice flux of $\dot{M}_{\rm{H}_{2}\rm{O}} = 3 \times 10^{-4} M_{\oplus} \: \rm{yr}^{-1}$.

We find that in this model no planetesimals form. We plot the solids surface density profiles at different time points in \fg{viscsigmasolids}. Because of the large value for $\alpha$, silicate particles interior to the snowline move with the gas velocity (the radial drift component to their velocity is negligible compared to the gas speed). Therefore, the solids surface density profile inside the snowline has the same power-law index as the gas profile. Outside the snowline, particles are fragmentation-limited, again due to the high $\alpha$-value. This leads to an $r^{-3/2}$ profile outside the snowline. Even further away from the star, particles are drift-limited and the solids profile evolves towards an $r^{-1}$ profile. If we would have chosen a smaller value for $\alpha$ in this time-dependent gas accretion model, such that planetesimals would form, viscous evolution would be slower and the position of the snowline would not change significantly during the planetesimal formation phase. Therefore, we conclude that viscous evolution does not play an important role during the planetesimal formation phase. Recently, \citet{2018A&A...614A..62D} concluded that for $\alpha \gtrsim 10^{-4}$, the build-up stage of the disk (when material is still falling onto the disk) is also not important for planetesimal formation. Of course, changes in the disk conditions due to viscous evolution would matter for the subsequent stages of planetesimal-planetesimal mergers and embryo migration, which occur after planetesimal formation and may take place on timescales longer than the viscous evolution timescale.

\subsection{Higher metallicity: icy and dry planetesimal formation}\label{sec:directSI}
\Fg{directSI} shows results for model run 1i, where we increased the metallicity $Z$ from 1\% to 2\%. Planetesimals form both outside and inside the snowline, but the icy planetesimals dominate the total mass in planetesimals. During the streaming instability phase outside the snowline, the pebble mass flux that reaches the snowline is equal to the critical pebble mass flux (\se{SIsnowline}), just as in the fiducial model. Before and after the icy planetesimal formation phase, however, the pebble mass flux delivered to the snowline is higher in this model than in the fiducial model because of the increased metallicity. Interior to the snowline the pebbles pile up after their ice has evaporated and they have fragmented to smaller sizes. The solids-to-gas ratio just interior to the snowline is not large enough to trigger streaming instability. However, as the pile up extends through the inner disk, dry planetesimals are formed closer to the star. This is because the surface density profile in the inner disk tends to an $r^{-3/2}$ power law: particle sizes are limited by fragmentation and the radial velocity is not dominated by the gas accretion velocity (as is the case in \fg{viscsigmasolids}, where $\alpha$ and hence the gas accretion velocity $v_{\rm{gas}}$ are higher). An $r^{-3/2}$ solids surface density power law is steeper than the gas surface density profile, leading to an increasing solids-to-gas ratio with decreasing distance to the star. The result is dry planetesimal formation that peaks close to the star. This effect was also described in \citet{2016A&A...594A.105D}.
Note that in our model, turbulent radial diffusion of particles is not accounted for (though it is implicitly included in our treatment of planetesimal formation outside the snowline; see \se{raddif}), and might oppose a particle pile up in the inner disk. However, small particles in the inner disk move inward with the gas accretion velocity, and therefore small particles are removed from (and delivered to) the inner pile-up region on a timescale that is of the same order as the radial diffusion timescale (because $v_{\rm{gas}} \sim \nu / r$; \eq{nu}).

In the work of \citet{2017A&A...608A..92D} no dry planetesimals were formed interior to the snowline even at high metallicities. This difference with our results is due to the fact that the dimensionless stopping times $\tau$ of particles involved in dry planetesimal formation in our simulations are $\sim$$10^{-3}$, and \citet{2017A&A...608A..92D} did not allow for streaming instability by particles with stopping times smaller than $10^{-2}$.

The ratio of the total amount of icy planetesimals that forms in model 1i ($Z = 0.02$) compared to the amount that forms in model 1a ($Z=0.01$) is much more than a factor two. This is because in model 1a, the pebble mass flux does not exceed the pebble mass flux threshold by a lot (see \fg{massflux_pltsmls}). Therefore, if we increase $Z$ by a factor two, the `excess' pebble mass flux becomes larger by a factor much larger than two.  

In model runs 1j and 1k we increase the `direct' planetesimal formation efficiency $\epsilon_{\rm{SI}}$ to 0.1 and 1.0, respectively (see \se{direct}). We find that the amount of dry planetesimals that forms close to the star increases by a factor $\sim$$3$ between model 1i and 1k, suggesting that the planetesimal formation efficiency is mildly dependent on the streaming instability efficiency.

\begin{figure}[t]
	\centering
		\includegraphics[width=0.49\textwidth]{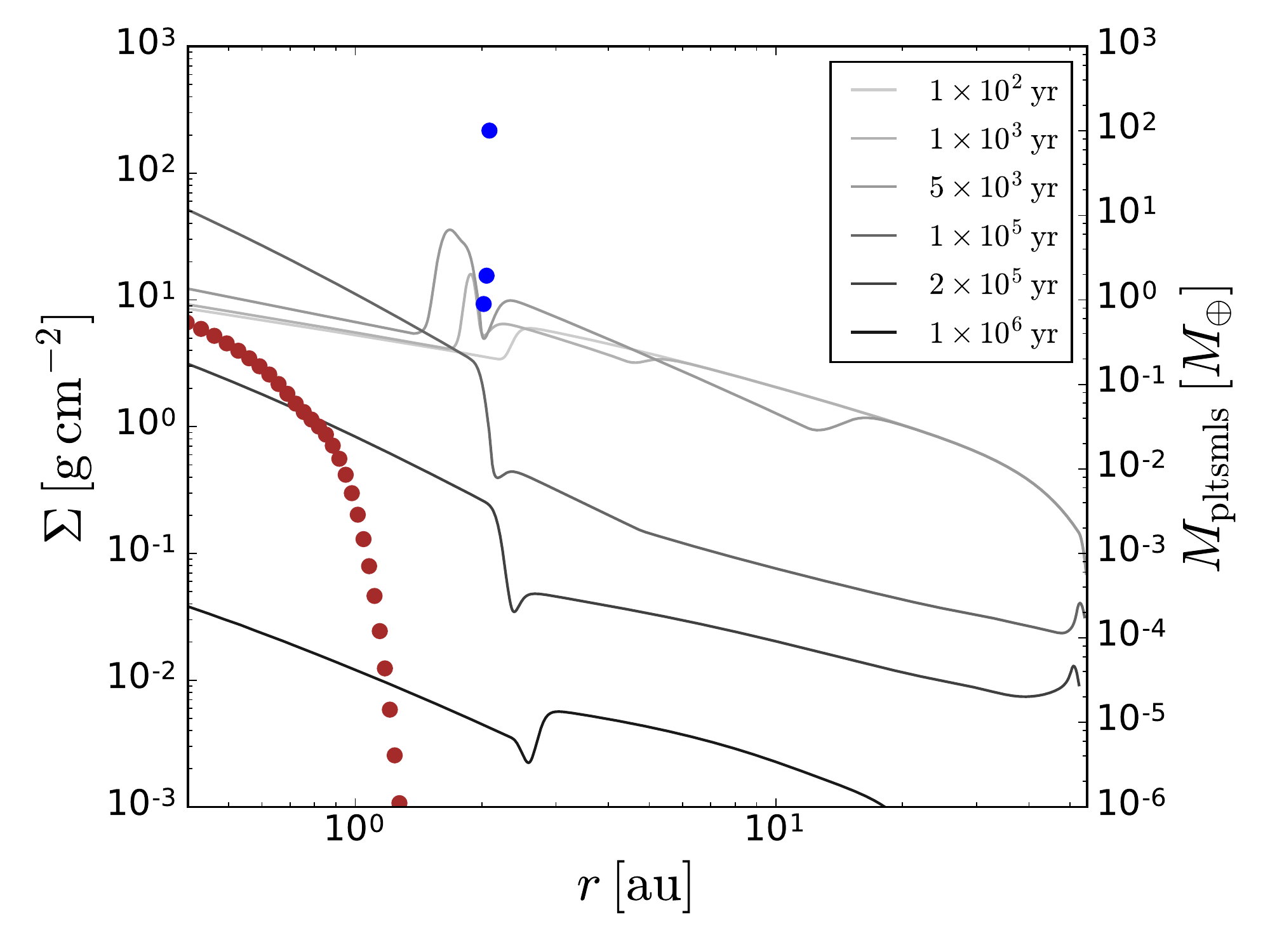}
\caption{Results for model 1i where $Z = 0.02$ and planetesimal formation occurs not only outside the snowline, but also interior to it. The masses of the icy planetesimal super-particles are plotted in blue; the masses of the water-poor planetesimal super-particles are plotted in brown. A massive annulus of planetesimals forms outside the snowline. Interior to the snowline, a traffic-jam effect leads to a high solids-to-gas ratio in the inner disk such that streaming instability is triggered by dry pebbles as well.\label{fig:directSI}}
\end{figure}

\subsection{Including pebble accretion}\label{sec:pa}
When small planetesimals have merged to larger bodies --- a process that we do not model in this work --- they can start to grow more efficiently by accreting pebbles \citep{2016A&A...586A..66V}. In our current models, we do not have information about the individual planetesimal sizes, because for simplicity we treat planetesimals as super-particle `sinks'  that are only characterised by a position and a total mass. However, if the planetesimals are residing in a narrow annulus, we can assume that the pebble accretion efficiency $\epsilon_{\rm{PA}}$ is proportional to the total mass in planetesimals, without having any information on the individual planetesimals. This assumption is only valid in the three-dimensional regime \citep{2017ASSL..445..197O}, when the planetesimals are residing in a vertical layer that is less extended than that of the pebbles, and only when planetesimal eccentricities and inclinations are small. The three-dimensional pebble accretion rate is given by \citep{2018arXiv180306149L}:
\begin{equation}\label{eq:pa}
\frac{d M_{\rm{pl, PA}}}{d t} = \epsilon_{\rm{PA}} \dot{M}_{\rm{peb}} =  \rm{max} \left(1, \frac{{\it A}_{3} {\it q_{p}} {\it r}}{\eta {\it H}_{\rm{peb}}}\right)  \dot{\it{M}}_{\rm{peb}}
\end{equation}
where $A_{3} = 0.39$ is a numerical constant \citep{2018arXiv180306150O}, $q_{p}$ is the mass ratio between the planetesimals and the star, and $\dot{M}_{\rm{peb}}$ is the pebble flux just outside the planetesimal annulus. 

We take our fiducial model, in which a narrow annulus of planetesimals forms outside the snowline (\se{fid}), and now take pebble accretion into account. We assume that the pebble mass flux available to streaming instability outside the snowline $\dot{M}_{\rm{peb, SI}}$ equals the mass flux that remains after the pebbles that are accreted by the already existing planetesimals have been removed: 
\begin{equation}
\dot{M}_{\rm{peb, SI}} = \dot{M}_{\rm{peb}} (1 - \epsilon_{\rm{PA}})
\end{equation}
The total mass loss rate of pebble super-particles due to pebble accretion is the negative of \eq{pa} and the individual pebble super-particle mass loss rates are calculated according to \eq{pebmassloss}.

The resulting pebble mass fluxes at the snowline and outside the planetesimal annulus, as well as the total mass in planetesimals, are plotted as a function of time in \fg{PA}. The planetesimal formation phase, which is characterised by the pebble mass flux that reaches the snowline being constant, is much shorter than in the fiducial model (\fg{massflux_pltsmls}). This is because pebble accretion reduces the pebble mass flux available for streaming instability $\dot{M}_{\rm{peb, SI}}$, which quickly becomes smaller than the critical pebble mass flux $\dot{M}_{\rm{peb, crit}}$. Therefore, even though the pebble mass flux delivered from the outer disk is still considerably larger than $\dot{M}_{\rm{peb, crit}}$, planetesimal formation stops and pebble accretion takes over. Planetesimal formation is a self-limiting process: the more planetesimals are formed, the larger the pebble accretion efficiency becomes, which eventually prohibits the formation of new planetesimals. The final total mass in planetesimals is 89.9 Earth masses, which is much larger than in the case without pebble accretion.

Two key effects will alter these results. First, planetesimals can self-excite each other to highly eccentric and inclined orbits when the individual planetesimals are large \citep{2015Natur.524..322L}, or when large embryos emerge as part of a runaway growth process \citep{1998Icar..131..171K}. In that case pebble accretion on the smaller planetesimals will terminate, because the relative motions between pebbles and planetesimals become too large \citep{2018arXiv180306149L}. Second, the eventual planetary embryos that emerge as the result of the combined planetesimal and pebble accretion are likely to migrate out of their birth zone due to \hbox{Type-I} migration \citep{2002ApJ...565.1257T}. Both effects will suppress the amount of planetesimals formed compared to the above estimate. A thorough analysis of the efficacy of this process is underway (Liu, Ormel, \& Johansen, submitted). Another effect of a large embryo crossing the snowline after the planetesimal formation phase has terminated could be that the pebble mass flux reaching the snowline becomes large enough for streaming instability again, such that a new generation of planetesimals forms.

\begin{figure}[t]
	\centering
		\includegraphics[width=0.49\textwidth]{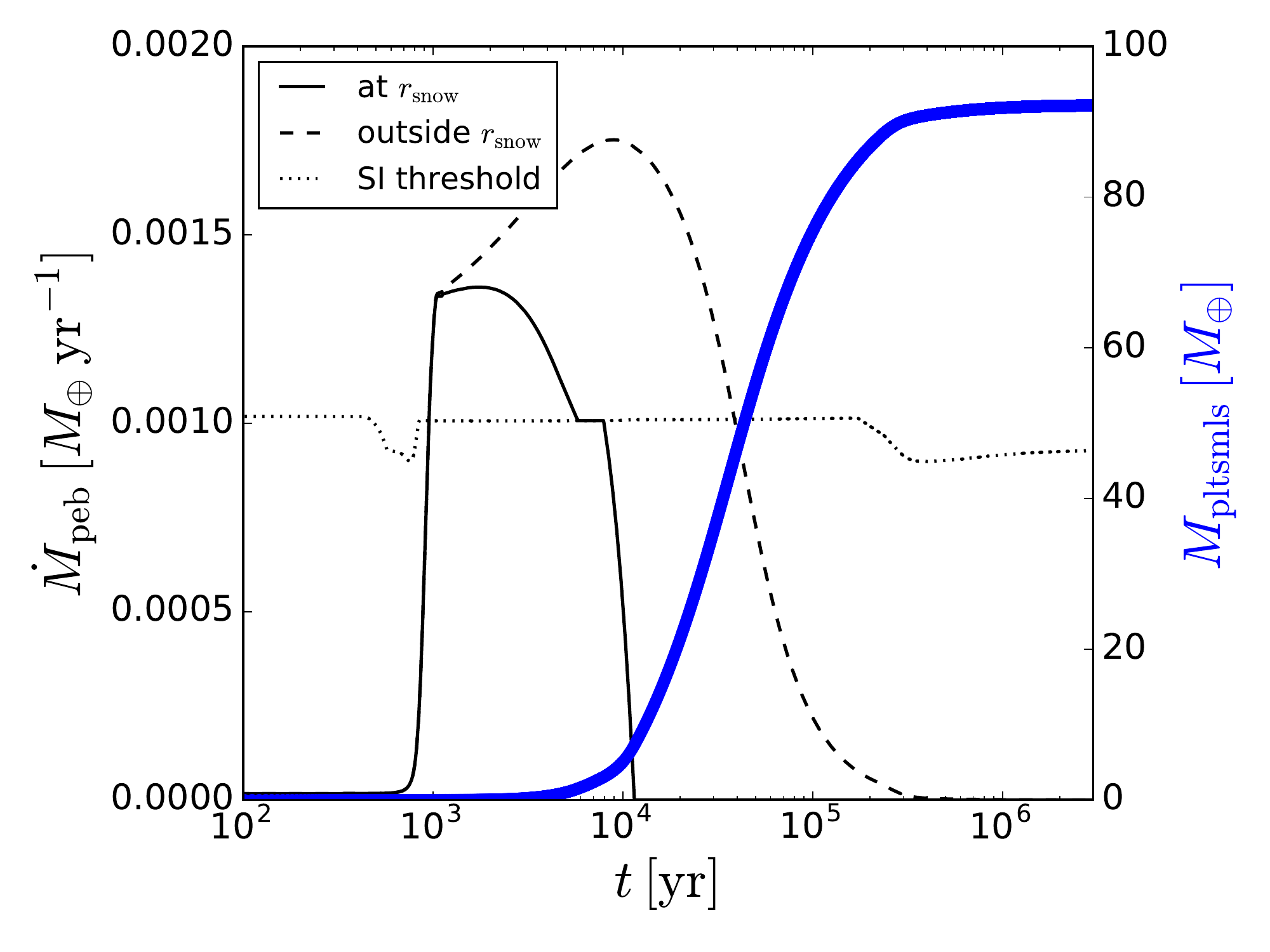}
\caption{Results for model 1h, which is the fiducial model including pebble accretion. The pebble mass flux arriving at the snowline is plotted by the solid black line; the pebble mass flux outside the planetesimal annulus is plotted by the dashed black line. The dotted black line corresponds to the streaming instability threshold pebble mass flux. The total mass in planetesimals including pebble accretion is given by the blue line. A little after $10^{4} \: \rm{yr}$, the pebble accretion efficiency is 100\% and no pebbles reach the snowline anymore.\label{fig:PA}}
\end{figure}

\subsection{Increasing the fragmentation threshold for icy particles}\label{sec:driftlim}
Up to now, we have fixed the fragmentation threshold velocity for icy particles at $v_{\rm{frag, icy}} = 10 \: \rm{m} \: \rm{s}^{-1}$. However, some molecular dynamics simulations suggest icy particles fragment only at higher velocities \citep{1997ApJ...480..647D,2009ApJ...702.1490W,2013A&A...559A..62W}. We therefore explore if and how the results change if we increase the icy fragmentation threshold from $10 \: \rm{m} \: \rm{s}^{-1}$ to $60 \: \rm{m} \: \rm{s}^{-1}$. Because of the higher fragmentation threshold, the sizes of particles outside the snowline are now limited by radial drift rather than fragmentation. In \fg{mf_1a_driftlim} the pebble mass fluxes $\dot{M}_{\rm{peb}}$ outside the snowline (dashed black line) and at the snowline (solid black line), as well as the critical pebble mass flux required for triggering streaming instability, are plotted as a function of time. First, as the stopping time of pebbles outside the snowline increases due to particle growth, the pebble mass flux increases and the critical pebble mass flux (which depends on the stopping time) increases as well. In this case particles outside the snowline reach dimensionless stopping times $\tau \sim 0.3$ at about $10^{3} \: \rm{yr}$, after which the critical pebble mass flux decreases. At around $6 \times 10^{3} \: \rm{yr}$ the stopping time at the snowline has reached a maximum of about $0.6$, after which it gradually become smaller again because of depletion of solids mass outside the snowline due to radial drift (see also \citet{2014A&A...572A.107L}), and the threshold for streaming instability on the pebble mass flux increases again. The total mass in planetesimals as a function of time is plotted in blue. The final total planetesimal mass is 17.1 Earth masses, which is larger than the planetesimals yield of the fiducial model. This is caused by the fact that the pebble mass flux outside the snowline reaches larger values in this model than in the fiducial model due to the higher stopping times, as well as by the fact that the pebble mass flux required for streaming instability is for a large period of time reduced compared to the fiducial model, again because of the higher stopping times. All in all this leads to more excess pebble mass flux outside the snowline that can be converted to planetesimals. Note also that the shape of the pebble mass flux as a function of time is different in the drift-limited case (\fg{mf_1a_driftlim}) than in the fragmentation-limited case (\fg{massflux_pltsmls}).

\begin{figure}[t]
	\centering
		\includegraphics[width=0.49\textwidth]{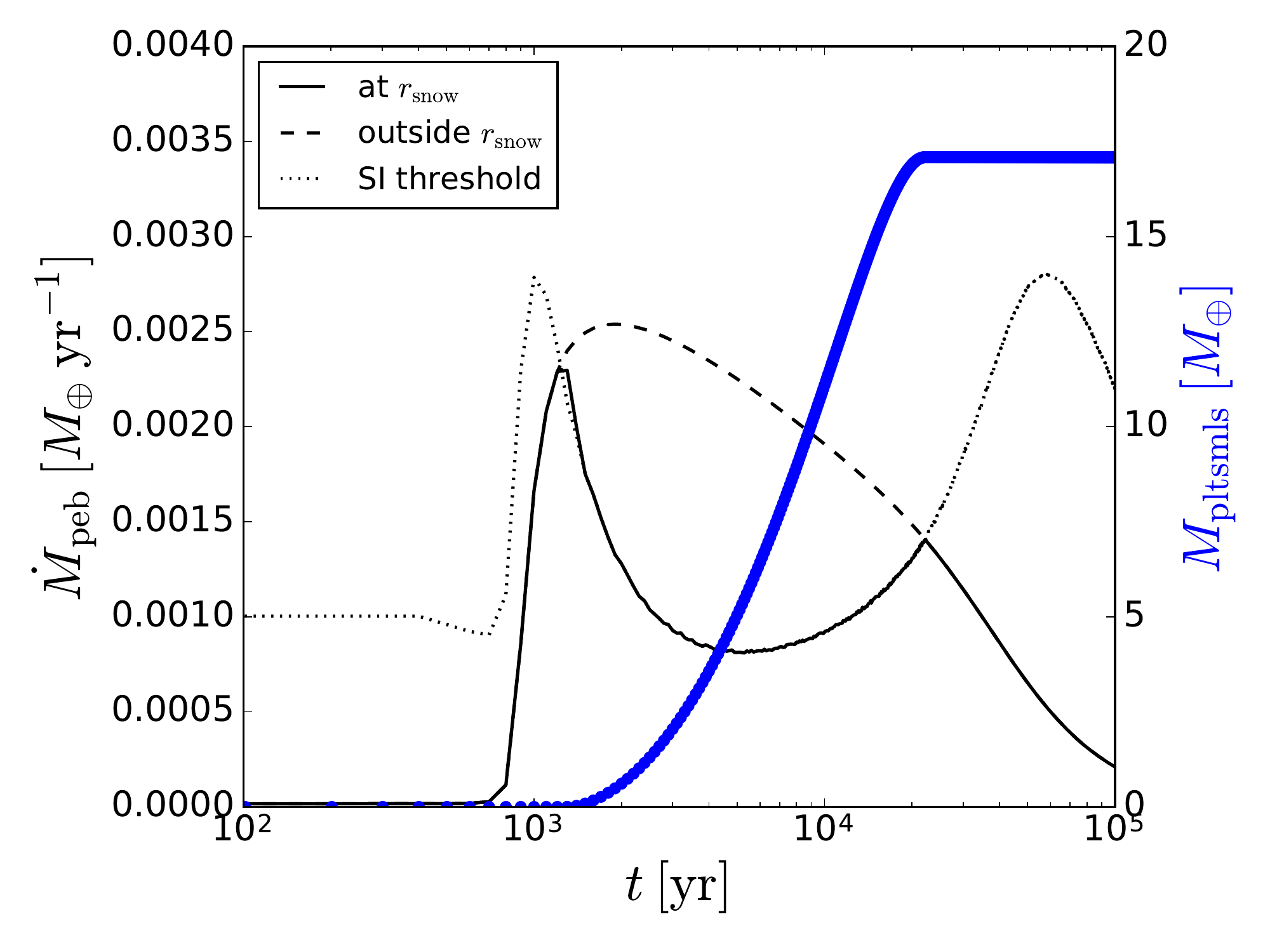}
\caption{Results for model 1l, which is the fiducial model but with icy fragmentation threshold $v_{\rm{frag, icy}}= 60 \: \rm{m} \: \rm{s}^{-1}$. In solid black: the pebble mass flux $\dot{M}_{\rm{peb}}$ that reaches the snowline; in dashed black: the pebble mass flux outside the snowline that is available for planetesimal formation outside the snowline; and in dotted black the critical pebble mass flux needed for streaming instability outside the snowline, all as a function of time. The total mass in planetesimals is plotted in blue.\label{fig:mf_1a_driftlim}}
\end{figure}

\section{Discussion}\label{sec:discussion}

We find that lower-mass stars are more efficient at forming planetesimals than higher-mass stars, {\it i.e.} disks around low-mass stars tend to convert a larger fraction of their initial solids content to planetesimals. This is because in our model, planetesimal formation works by virtue of a high flux of pebbles through the disk. The pebbles-to-gas mass flux ratio, which dictates planetesimal formation outside the snowline, depends inversely on snowline location, which for lower-mass stars is closer in than for higher-mass stars that have hotter disks. The pebbles-to-planetesimals conversion efficiency also depends on other disk parameters that may vary across the stellar mass range, however, such as the outer disk radius, the turbulence parameter $\alpha$, and the metallicity. A spread in disk properties ({\it e.g.}, \citet{2017AJ....153..240A,2017ApJ...847...31M,2017A&A...604A.127M,2017A&A...606A..88T,2018ApJ...859...21A}) might obscure any observable correlation between stellar mass and planetesimal formation efficiency.

For a given stellar mass and gas accretion rate, the total amount of planetesimals that forms depends sensitively on the value of the turbulence parameter $\alpha$. This is partly due to the nature of the viscous gas disk model: for a given gas accretion rate, the gas surface density is larger for lower $\alpha$. In our model this means that for a given stellar mass and metallicity, a lower $\alpha$ leads to a larger pebble mass flux. This conclusion was already made in \citet{SO2017}. Another quantity that proved to be important in our models is the fragmentation velocity threshold, because it determines the drift velocity of pebbles and hence the pebble flux. Therefore, more stringent laboratory constraints on the fragmentation threshold in the snowline region (including snowline-specific effects such as sintering), as well as constraints on particle sizes around observed snowlines (such as around FU Ori objects \citep{2016Natur.535..258C,2015ApJ...815L..15B,2017A&A...605L...2S})), would help narrow down the parameter space for our models. Additionally, recent studies predict observable signatures of drifting pebbles in the gas-phase abundances of $\rm{CO}$ and $\rm{CO}_{2}$ \citep{2017MNRAS.469.3994B,2018A&A...611A..80B,2018arXiv180801840K}, which could constrain the pebble mass flux observationally.

Because the crucial factor in our planetesimal formation model is a high pebble flux, planetesimals form in the early stages of disk evolution (in the first $\sim$$10^{5}$ years) when the pebble mass flux is still high. A side effect of fast planetesimal formation is the well-known `radial drift' problem: depletion of the solids content of the disk on short timescales is in disagreement with observations, which show that small solid particles are present in the outer regions of protoplanetary disks at later ages. 
A solution to this problem could be to take into account a particle size distribution. In this work, we assume a mono-disperse particle size distribution at each point in space and time, and thereby only focus on the large particles. Large particles dominate the mass, but small particles that do not drift significantly might be present at all times. This could also be the clue to late planetesimal formation: while in our model only early-stage planetesimal formation is possible due to the depletion of solids over time, other works have suggested models in which late planetesimal formation occurs in the outer disk at a late evolutionary disk stage due to gas depletion \citep{2017ApJ...839...16C}. Also, the early formation of a protoplanet could halt radial drift \citep{2012ApJ...756...70K, 2016Icar..267..368M}.
Finally, another possible solution to the radial drift problem in the context of our model is choosing a larger outer disk radius $r_{\rm{out}}$, which leads to longer drift timescales in the outer regions of the disk.

The Lagrangian smooth-particle model presented in this paper can be used for different research directions that we have not explored yet in this paper. Characteristics of individual (groups of) particles can be naturally followed during their evolution in time and through the disk. In this work we have focused on the water content, but any other compositional information could be included as well. For example, the D/H ratio of water depends on the local conditions in the evolving protoplanetary disk \citep{2013Icar..226..256Y}, and therefore measurements of the D/H content of bodies in the Solar System might provide information on where they formed \citep{2015Sci...350..795H, 2017RSPTA.37550390H}. An interesting direction for further research using our model would therefore be to focus on the D/H ratio of water in planetesimals by keeping track of where and when in the disk their water is incorporated.
One could also think of different disk structures than the ones we have considered in this paper. Axisymmetric features such as rings and gaps seem to be ubiquitous in the gas and dust of observed disks ({\it e.g.}, \citet{2015ApJ...808L...3A,2015ApJ...802L..17A,2016ApJ...820L..40A,2016PhRvL.117y1101I,2016ApJ...819L...7N,2018A&A...610A..24F}). In our model scenario, planetesimal formation occurs at earlier stages than the ages of the observed disks. However, in principle our model could deal with any disk structure. It would be interesting to investigate what the consequences would be for planetesimal formation if one assumes a disk with radial pressure bumps rather than a smooth disk as we did in the current work. Such structures could also alleviate the radial drift problem \citep{2012A&A...545A..81P}.

The formation of planetesimals from dust is clearly only the first piece of the planet formation puzzle. The next step is how planetesimals grow into protoplanets and beyond, for which we need an N-body model that follows the interactions between planetesimals, their growth by mutual collisions and pebble accretion, and their migration through the disk. The combination of the Lagrangian planetesimal formation model presented in this paper with such an N-body code will allow us to model the formation of entire planetary systems from A to Z while automatically following composition, which is a research direction we are planning to pursue. Our first target is the $\rm{H}_{2}\rm{O}$ fraction of the TRAPPIST-1 system (Schoonenberg et al., in preparation).

\section{Conclusions}\label{sec:conclusions}
Our main findings can be summarised as follows:

\begin{enumerate}
    \item Planetesimals form early, when the pebble mass flux is still high. In the context of our model, in this early formation phase the migration of the snowline due to a decreasing gas accretion rate is not important.
    \item Planetesimals form preferentially in a narrow annulus outside the snowline. Even if rocky planetesimals are formed interior to the snowline, icy planetesimals dominate the total planetesimal mass.
    \item Pebble accretion leads to self-limiting planetesimal formation (the more planetesimals form, the higher the pebble accretion efficiency and the smaller the pebble mass flux available to form new planetesimals). However, taking into account migration and planetesimal-planetesimal scattering will change this picture.
    \item The planetesimal formation efficiency depends on the location of the water snowline. The cooler the disk (the closer-in the water snowline), the higher the pebbles-to-planetesimals conversion efficiency. Therefore, in general, we find that low-mass stars are better at producing planetesimals than high-mass stars. However, this result also depends on other factors such as the compactness of disks, and may change for gas disks that deviate from a steady-state $\alpha$-viscosity disk where the gas surface density depends on a constant value of $\alpha$ as we assumed in this work.
\end{enumerate}

\begin{acknowledgements}
    D.S.\ and C.W.O\ are supported by the Netherlands Organization for Scientific Research (NWO; VIDI project 639.042.422). S.K. acknowledges support from NASA through Hubble Fellowship grant HST-HF2-51394 awarded by the Space Telescope Science Institute, which is operated by the Association of Universities for Research in Astronomy, Inc., for NASA, under contract NAS5-26555. D.S. would like to thank Carsten Dominik for useful discussions and Shigeru Ida for pointing out a typographical error in \citet{SO2017}. We thank Beibei Liu, Joanna {Dr{\c a}{\.z}kowska}, and the anonymous referee for constructive feedback on the manuscript.\end{acknowledgements}

\bibliography{SPMbib}

\begin{thebibliography}{91}
\expandafter\ifx\csname natexlab\endcsname\relax\def\natexlab#1{#1}\fi

\bibitem[{{Akiyama} {et~al.}(2015){Akiyama}, {Muto}, {Kusakabe}, {Kataoka},
  {Hashimoto}, {Tsukagoshi}, {Kwon}, {Kudo}, {Kandori}, {Grady}, {Takami},
  {Janson}, {Kuzuhara}, {Henning}, {Sitko}, {Carson}, {Mayama}, {Currie},
  {Thalmann}, {Wisniewski}, {Momose}, {Ohashi}, {Abe}, {Brandner}, {Brandt},
  {Egner}, {Feldt}, {Goto}, {Guyon}, {Hayano}, {Hayashi}, {Hayashi}, {Hodapp},
  {Ishi}, {Iye}, {Knapp}, {Matsuo}, {Mcelwain}, {Miyama}, {Morino},
  {Moro-Martin}, {Nishimura}, {Pyo}, {Serabyn}, {Suenaga}, {Suto}, {Suzuki},
  {Takahashi}, {Takato}, {Terada}, {Tomono}, {Turner}, {Watanabe}, {Yamada},
  {Takami}, {Usuda}, \& {Tamura}}]{2015ApJ...802L..17A}
{Akiyama}, E., {Muto}, T., {Kusakabe}, N., {et~al.} 2015, \apjl, 802, L17

\bibitem[{{ALMA Partnership} {et~al.}(2015){ALMA Partnership}, {Brogan},
  {P{\'e}rez}, {Hunter}, {Dent}, {Hales}, {Hills}, {Corder}, {Fomalont},
  {Vlahakis}, {Asaki}, {Barkats}, {Hirota}, {Hodge}, {Impellizzeri}, {Kneissl},
  {Liuzzo}, {Lucas}, {Marcelino}, {Matsushita}, {Nakanishi}, {Phillips},
  {Richards}, {Toledo}, {Aladro}, {Broguiere}, {Cortes}, {Cortes}, {Espada},
  {Galarza}, {Garcia-Appadoo}, {Guzman-Ramirez}, {Humphreys}, {Jung}, {Kameno},
  {Laing}, {Leon}, {Marconi}, {Mignano}, {Nikolic}, {Nyman}, {Radiszcz},
  {Remijan}, {Rod{\'o}n}, {Sawada}, {Takahashi}, {Tilanus}, {Vila Vilaro},
  {Watson}, {Wiklind}, {Akiyama}, {Chapillon}, {de Gregorio-Monsalvo}, {Di
  Francesco}, {Gueth}, {Kawamura}, {Lee}, {Nguyen Luong}, {Mangum}, {Pietu},
  {Sanhueza}, {Saigo}, {Takakuwa}, {Ubach}, {van Kempen}, {Wootten},
  {Castro-Carrizo}, {Francke}, {Gallardo}, {Garcia}, {Gonzalez}, {Hill},
  {Kaminski}, {Kurono}, {Liu}, {Lopez}, {Morales}, {Plarre}, {Schieven},
  {Testi}, {Videla}, {Villard}, {Andreani}, {Hibbard}, \&
  {Tatematsu}}]{2015ApJ...808L...3A}
{ALMA Partnership}, {Brogan}, C.~L., {P{\'e}rez}, L.~M., {et~al.} 2015, \apjl,
  808, L3

\bibitem[{{Andrews} {et~al.}(2016){Andrews}, {Wilner}, {Zhu}, {Birnstiel},
  {Carpenter}, {P{\'e}rez}, {Bai}, {{\"O}berg}, {Hughes}, {Isella}, \&
  {Ricci}}]{2016ApJ...820L..40A}
{Andrews}, S.~M., {Wilner}, D.~J., {Zhu}, Z., {et~al.} 2016, \apjl, 820, L40

\bibitem[{{Ansdell} {et~al.}(2017){Ansdell}, {Williams}, {Manara}, {Miotello},
  {Facchini}, {van der Marel}, {Testi}, \& {van
  Dishoeck}}]{2017AJ....153..240A}
{Ansdell}, M., {Williams}, J.~P., {Manara}, C.~F., {et~al.} 2017, \aj, 153, 240

\bibitem[{{Ansdell} {et~al.}(2018){Ansdell}, {Williams}, {Trapman}, {van
  Terwisga}, {Facchini}, {Manara}, {van der Marel}, {Miotello}, {Tazzari},
  {Hogerheijde}, {Guidi}, {Testi}, \& {van Dishoeck}}]{2018ApJ...859...21A}
{Ansdell}, M., {Williams}, J.~P., {Trapman}, L., {et~al.} 2018, \apj, 859, 21

\bibitem[{{Armitage} {et~al.}(2016){Armitage}, {Eisner}, \&
  {Simon}}]{ArmitageEtal2016}
{Armitage}, P.~J., {Eisner}, J.~A., \& {Simon}, J.~B. 2016, \apjl, 828, L2

\bibitem[{{Bai} \& {Stone}(2010)}]{BaiStone2010i}
{Bai}, X.-N. \& {Stone}, J.~M. 2010, \apj, 722, 1437

\bibitem[{{Banzatti} {et~al.}(2015){Banzatti}, {Pinilla}, {Ricci},
  {Pontoppidan}, {Birnstiel}, \& {Ciesla}}]{2015ApJ...815L..15B}
{Banzatti}, A., {Pinilla}, P., {Ricci}, L., {et~al.} 2015, \apjl, 815, L15

\bibitem[{{Benz}(2000)}]{2000SSRv...92..279B}
{Benz}, W. 2000, \ssr, 92, 279

\bibitem[{{Birnstiel} {et~al.}(2012){Birnstiel}, {Klahr}, \&
  {Ercolano}}]{2012A&A...539A.148B}
{Birnstiel}, T., {Klahr}, H., \& {Ercolano}, B. 2012, \aap, 539, A148

\bibitem[{{Blum} \& {M{\"u}nch}(1993)}]{1993Icar..106..151B}
{Blum}, J. \& {M{\"u}nch}, M. 1993, \icarus, 106, 151

\bibitem[{{Blum} \& {Wurm}(2000)}]{2000Icar..143..138B}
{Blum}, J. \& {Wurm}, G. 2000, \icarus, 143, 138

\bibitem[{{Booth} {et~al.}(2017){Booth}, {Clarke}, {Madhusudhan}, \&
  {Ilee}}]{2017MNRAS.469.3994B}
{Booth}, R.~A., {Clarke}, C.~J., {Madhusudhan}, N., \& {Ilee}, J.~D. 2017,
  \mnras, 469, 3994

\bibitem[{{Bosman} {et~al.}(2018){Bosman}, {Tielens}, \& {van
  Dishoeck}}]{2018A&A...611A..80B}
{Bosman}, A.~D., {Tielens}, A.~G.~G.~M., \& {van Dishoeck}, E.~F. 2018, \aap,
  611, A80

\bibitem[{{Carrera} {et~al.}(2017){Carrera}, {Gorti}, {Johansen}, \&
  {Davies}}]{2017ApJ...839...16C}
{Carrera}, D., {Gorti}, U., {Johansen}, A., \& {Davies}, M.~B. 2017, \apj, 839,
  16

\bibitem[{{Carrera} {et~al.}(2015){Carrera}, {Johansen}, \&
  {Davies}}]{2015A&A...579A..43C}
{Carrera}, D., {Johansen}, A., \& {Davies}, M.~B. 2015, \aap, 579, A43

\bibitem[{Chapman \& Cowling(1970)}]{chapman1970mathematical}
Chapman, S. \& Cowling, T. 1970, The Mathematical Theory of Non-uniform Gases:
  An Account of the Kinetic Theory of Viscosity, Thermal Conduction and
  Diffusion in Gases, Cambridge Mathematical Library (Cambridge University
  Press)

\bibitem[{{Chokshi} {et~al.}(1993){Chokshi}, {Tielens}, \&
  {Hollenbach}}]{1993ApJ...407..806C}
{Chokshi}, A., {Tielens}, A.~G.~G.~M., \& {Hollenbach}, D. 1993, \apj, 407, 806

\bibitem[{{Cieza} {et~al.}(2016){Cieza}, {Casassus}, {Tobin}, {Bos},
  {Williams}, {Perez}, {Zhu}, {Caceres}, {Canovas}, {Dunham}, {Hales},
  {Prieto}, {Principe}, {Schreiber}, {Ruiz-Rodriguez}, \&
  {Zurlo}}]{2016Natur.535..258C}
{Cieza}, L.~A., {Casassus}, S., {Tobin}, J., {et~al.} 2016, \nat, 535, 258

\bibitem[{{Cuzzi} \& {Zahnle}(2004)}]{2004ApJ...614..490C}
{Cuzzi}, J.~N. \& {Zahnle}, K.~J. 2004, \apj, 614, 490

\bibitem[{{Dominik} \& {Tielens}(1997)}]{1997ApJ...480..647D}
{Dominik}, C. \& {Tielens}, A.~G.~G.~M. 1997, \apj, 480, 647

\bibitem[{{Dr{\c a}{\.z}kowska} \& {Alibert}(2017)}]{2017A&A...608A..92D}
{Dr{\c a}{\.z}kowska}, J. \& {Alibert}, Y. 2017, \aap, 608, A92

\bibitem[{{Dr{\c a}{\.z}kowska} {et~al.}(2016){Dr{\c a}{\.z}kowska}, {Alibert},
  \& {Moore}}]{2016A&A...594A.105D}
{Dr{\c a}{\.z}kowska}, J., {Alibert}, Y., \& {Moore}, B. 2016, \aap, 594, A105

\bibitem[{{Dr{\c a}{\.z}kowska} \& {Dullemond}(2014)}]{DrazkowskaDullemond2014}
{Dr{\c a}{\.z}kowska}, J. \& {Dullemond}, C.~P. 2014, \aap, 572, A78

\bibitem[{{Dr{\c a}{\.z}kowska} \& {Dullemond}(2018)}]{2018A&A...614A..62D}
{Dr{\c a}{\.z}kowska}, J. \& {Dullemond}, C.~P. 2018, \aap, 614, A62

\bibitem[{{Fedele} {et~al.}(2018){Fedele}, {Tazzari}, {Booth}, {Testi},
  {Clarke}, {Pascucci}, {Kospal}, {Semenov}, {Bruderer}, {Henning}, \&
  {Teague}}]{2018A&A...610A..24F}
{Fedele}, D., {Tazzari}, M., {Booth}, R., {et~al.} 2018, \aap, 610, A24

\bibitem[{{Frank} {et~al.}(2002){Frank}, {King}, \&
  {Raine}}]{2002apa..book.....F}
{Frank}, J., {King}, A., \& {Raine}, D.~J. 2002, {Accretion Power in
  Astrophysics: Third Edition}, 398

\bibitem[{{Garaud} \& {Lin}(2007)}]{2007ApJ...654..606G}
{Garaud}, P. \& {Lin}, D.~N.~C. 2007, \apj, 654, 606

\bibitem[{{Gonzalez} {et~al.}(2017){Gonzalez}, {Laibe}, \&
  {Maddison}}]{2017MNRAS.467.1984G}
{Gonzalez}, J.-F., {Laibe}, G., \& {Maddison}, S.~T. 2017, \mnras, 467, 1984

\bibitem[{{Gundlach} \& {Blum}(2015)}]{2015ApJ...798...34G}
{Gundlach}, B. \& {Blum}, J. 2015, \apj, 798, 34

\bibitem[{{Hallis}(2017)}]{2017RSPTA.37550390H}
{Hallis}, L.~J. 2017, Philosophical Transactions of the Royal Society of London
  Series A, 375, 20150390

\bibitem[{{Hallis} {et~al.}(2015){Hallis}, {Huss}, {Nagashima}, {Taylor},
  {Halld{\'o}rsson}, {Hilton}, {Mottl}, \& {Meech}}]{2015Sci...350..795H}
{Hallis}, L.~J., {Huss}, G.~R., {Nagashima}, K., {et~al.} 2015, Science, 350,
  795

\bibitem[{{Hartmann} {et~al.}(1998){Hartmann}, {Calvet}, {Gullbring}, \&
  {D'Alessio}}]{1998ApJ...495..385H}
{Hartmann}, L., {Calvet}, N., {Gullbring}, E., \& {D'Alessio}, P. 1998, \apj,
  495, 385

\bibitem[{{Ida} \& {Guillot}(2016)}]{2016A&A...596L...3I}
{Ida}, S. \& {Guillot}, T. 2016, \aap, 596, L3

\bibitem[{{Isella} {et~al.}(2016){Isella}, {Guidi}, {Testi}, {Liu}, {Li}, {Li},
  {Weaver}, {Boehler}, {Carperter}, {De Gregorio-Monsalvo}, {Manara}, {Natta},
  {P{\'e}rez}, {Ricci}, {Sargent}, {Tazzari}, \&
  {Turner}}]{2016PhRvL.117y1101I}
{Isella}, A., {Guidi}, G., {Testi}, L., {et~al.} 2016, Physical Review Letters,
  117, 251101

\bibitem[{{Johansen} {et~al.}(2007){Johansen}, {Oishi}, {Mac Low}, {Klahr},
  {Henning}, \& {Youdin}}]{2007Natur.448.1022J}
{Johansen}, A., {Oishi}, J.~S., {Mac Low}, M.-M., {et~al.} 2007, \nat, 448,
  1022

\bibitem[{{Johansen} \& {Youdin}(2007)}]{2007ApJ...662..627J}
{Johansen}, A. \& {Youdin}, A. 2007, \apj, 662, 627

\bibitem[{{Johansen} {et~al.}(2009){Johansen}, {Youdin}, \& {Mac
  Low}}]{2009ApJ...704L..75J}
{Johansen}, A., {Youdin}, A., \& {Mac Low}, M.-M. 2009, \apjl, 704, L75

\bibitem[{{Kenyon} \& {Hartmann}(1987)}]{1987ApJ...323..714K}
{Kenyon}, S.~J. \& {Hartmann}, L. 1987, \apj, 323, 714

\bibitem[{{Kobayashi} {et~al.}(2012){Kobayashi}, {Ormel}, \&
  {Ida}}]{2012ApJ...756...70K}
{Kobayashi}, H., {Ormel}, C.~W., \& {Ida}, S. 2012, \apj, 756, 70

\bibitem[{{Kokubo} \& {Ida}(1998)}]{1998Icar..131..171K}
{Kokubo}, E. \& {Ida}, S. 1998, \icarus, 131, 171

\bibitem[{{Krijt} {et~al.}(2016){Krijt}, {Ormel}, {Dominik}, \&
  {Tielens}}]{2016A&A...586A..20K}
{Krijt}, S., {Ormel}, C.~W., {Dominik}, C., \& {Tielens}, A.~G.~G.~M. 2016,
  \aap, 586, A20

\bibitem[{{Krijt} {et~al.}(2018){Krijt}, {Schwarz}, {Bergin}, \&
  {Ciesla}}]{2018arXiv180801840K}
{Krijt}, S., {Schwarz}, K.~R., {Bergin}, E.~A., \& {Ciesla}, F.~J. 2018, ArXiv
  e-prints [\eprint[arXiv]{1808.01840}]

\bibitem[{L.~Hicks \& Liebrock(2000)}]{hicks}
L.~Hicks, D. \& Liebrock, L. 2000, 112, 63

\bibitem[{{Laibe} {et~al.}(2008){Laibe}, {Gonzalez}, {Fouchet}, \&
  {Maddison}}]{2008A&A...487..265L}
{Laibe}, G., {Gonzalez}, J.-F., {Fouchet}, L., \& {Maddison}, S.~T. 2008, \aap,
  487, 265

\bibitem[{{Lambrechts} \& {Johansen}(2014)}]{2014A&A...572A.107L}
{Lambrechts}, M. \& {Johansen}, A. 2014, \aap, 572, A107

\bibitem[{{Levison} {et~al.}(2015){Levison}, {Kretke}, \&
  {Duncan}}]{2015Natur.524..322L}
{Levison}, H.~F., {Kretke}, K.~A., \& {Duncan}, M.~J. 2015, \nat, 524, 322

\bibitem[{{Lichtenegger} \& {Komle}(1991)}]{1991Icar...90..319L}
{Lichtenegger}, H.~I.~M. \& {Komle}, N.~I. 1991, \icarus, 90, 319

\bibitem[{{Liu} \& {Ormel}(2018)}]{2018arXiv180306149L}
{Liu}, B. \& {Ormel}, C.~W. 2018, ArXiv e-prints [\eprint[arXiv]{1803.06149}]

\bibitem[{Liu \& Liu(2003)}]{liu2003smoothed}
Liu, G. \& Liu, M. 2003, Smoothed Particle Hydrodynamics: A Meshfree Particle
  Method (World Scientific)

\bibitem[{{Lodders}(2003)}]{2003ApJ...591.1220L}
{Lodders}, K. 2003, \apj, 591, 1220

\bibitem[{{Lorek} {et~al.}(2016){Lorek}, {Gundlach}, {Lacerda}, \&
  {Blum}}]{2016A&A...587A.128L}
{Lorek}, S., {Gundlach}, B., {Lacerda}, P., \& {Blum}, J. 2016, \aap, 587, A128

\bibitem[{{Lynden-Bell} \& {Pringle}(1974)}]{1974MNRAS.168..603L}
{Lynden-Bell}, D. \& {Pringle}, J.~E. 1974, \mnras, 168, 603

\bibitem[{{Manara} {et~al.}(2017){Manara}, {Testi}, {Herczeg}, {Pascucci},
  {Alcal{\'a}}, {Natta}, {Antoniucci}, {Fedele}, {Mulders}, {Henning},
  {Mohanty}, {Prusti}, \& {Rigliaco}}]{2017A&A...604A.127M}
{Manara}, C.~F., {Testi}, L., {Herczeg}, G.~J., {et~al.} 2017, \aap, 604, A127

\bibitem[{{Martin} \& {Livio}(2012)}]{2012MNRAS.425L...6M}
{Martin}, R.~G. \& {Livio}, M. 2012, \mnras, 425, L6

\bibitem[{{Morbidelli} {et~al.}(2016){Morbidelli}, {Bitsch}, {Crida},
  {Gounelle}, {Guillot}, {Jacobson}, {Johansen}, {Lambrechts}, \&
  {Lega}}]{2016Icar..267..368M}
{Morbidelli}, A., {Bitsch}, B., {Crida}, A., {et~al.} 2016, \icarus, 267, 368

\bibitem[{{Morbidelli} {et~al.}(2015){Morbidelli}, {Lambrechts}, {Jacobson}, \&
  {Bitsch}}]{2015Icar..258..418M}
{Morbidelli}, A., {Lambrechts}, M., {Jacobson}, S., \& {Bitsch}, B. 2015,
  \icarus, 258, 418

\bibitem[{{Mulders}(2018)}]{2018arXiv180500023M}
{Mulders}, G.~D. 2018, ArXiv e-prints [\eprint[arXiv]{1805.00023}]

\bibitem[{{Mulders} {et~al.}(2015){Mulders}, {Pascucci}, \&
  {Apai}}]{2015ApJ...814..130M}
{Mulders}, G.~D., {Pascucci}, I., \& {Apai}, D. 2015, \apj, 814, 130

\bibitem[{{Mulders} {et~al.}(2017){Mulders}, {Pascucci}, {Manara}, {Testi},
  {Herczeg}, {Henning}, {Mohanty}, \& {Lodato}}]{2017ApJ...847...31M}
{Mulders}, G.~D., {Pascucci}, I., {Manara}, C.~F., {et~al.} 2017, \apj, 847, 31

\bibitem[{{Nakagawa} {et~al.}(1986){Nakagawa}, {Sekiya}, \&
  {Hayashi}}]{NakagawaEtal1986}
{Nakagawa}, Y., {Sekiya}, M., \& {Hayashi}, C. 1986, Icarus, 67, 375

\bibitem[{{Nomura} {et~al.}(2016){Nomura}, {Tsukagoshi}, {Kawabe}, {Ishimoto},
  {Okuzumi}, {Muto}, {Kanagawa}, {Ida}, {Walsh}, {Millar}, \&
  {Bai}}]{2016ApJ...819L...7N}
{Nomura}, H., {Tsukagoshi}, T., {Kawabe}, R., {et~al.} 2016, \apjl, 819, L7

\bibitem[{{Oka} {et~al.}(2011){Oka}, {Nakamoto}, \&
  {Ida}}]{2011ApJ...738..141O}
{Oka}, A., {Nakamoto}, T., \& {Ida}, S. 2011, \apj, 738, 141

\bibitem[{{Okuzumi} {et~al.}(2012){Okuzumi}, {Tanaka}, {Kobayashi}, \&
  {Wada}}]{2012ApJ...752..106O}
{Okuzumi}, S., {Tanaka}, H., {Kobayashi}, H., \& {Wada}, K. 2012, \apj, 752,
  106

\bibitem[{{Ormel}(2017)}]{2017ASSL..445..197O}
{Ormel}, C.~W. 2017, in Astrophysics and Space Science Library, Vol. 445,
  Astrophysics and Space Science Library, ed. M.~{Pessah} \& O.~{Gressel}, 197

\bibitem[{{Ormel} \& {Cuzzi}(2007)}]{2007A&A...466..413O}
{Ormel}, C.~W. \& {Cuzzi}, J.~N. 2007, \aap, 466, 413

\bibitem[{{Ormel} \& {Liu}(2018)}]{2018arXiv180306150O}
{Ormel}, C.~W. \& {Liu}, B. 2018, ArXiv e-prints [\eprint[arXiv]{1803.06150}]

\bibitem[{{Ormel} {et~al.}(2017){Ormel}, {Liu}, \& {Schoonenberg}}]{OLS2017}
{Ormel}, C.~W., {Liu}, B., \& {Schoonenberg}, D. 2017, \aap, 604, A1

\bibitem[{{Pinilla} {et~al.}(2012){Pinilla}, {Benisty}, \&
  {Birnstiel}}]{2012A&A...545A..81P}
{Pinilla}, P., {Benisty}, M., \& {Birnstiel}, T. 2012, \aap, 545, A81

\bibitem[{{Pollack} {et~al.}(1996){Pollack}, {Hubickyj}, {Bodenheimer},
  {Lissauer}, {Podolak}, \& {Greenzweig}}]{PollackEtal1996}
{Pollack}, J.~B., {Hubickyj}, O., {Bodenheimer}, P., {et~al.} 1996, Icarus,
  124, 62

\bibitem[{{Ros} \& {Johansen}(2013)}]{2013A&A...552A.137R}
{Ros}, K. \& {Johansen}, A. 2013, \aap, 552, A137

\bibitem[{{Safronov}(1969)}]{Safronov1969}
{Safronov}, V.~S. 1969, {Evolution of the Protoplanetary Cloud and Formation of
  Earth and the Planets}, ed. V.~S. Safronov (Moscow: Nauka. Transl. 1972 NASA
  Tech. F-677)

\bibitem[{{Saito} \& {Sirono}(2011)}]{2011ApJ...728...20S}
{Saito}, E. \& {Sirono}, S.-i. 2011, \apj, 728, 20

\bibitem[{{Sato} {et~al.}(2016){Sato}, {Okuzumi}, \& {Ida}}]{SatoEtal2016}
{Sato}, T., {Okuzumi}, S., \& {Ida}, S. 2016, \aap, 589, A15

\bibitem[{{Schoonenberg} {et~al.}(2017){Schoonenberg}, {Okuzumi}, \&
  {Ormel}}]{2017A&A...605L...2S}
{Schoonenberg}, D., {Okuzumi}, S., \& {Ormel}, C.~W. 2017, \aap, 605, L2

\bibitem[{{Schoonenberg} \& {Ormel}(2017)}]{SO2017}
{Schoonenberg}, D. \& {Ormel}, C.~W. 2017, \aap, 602, A21

\bibitem[{{Shakura} \& {Sunyaev}(1973)}]{1973A&A....24..337S}
{Shakura}, N.~I. \& {Sunyaev}, R.~A. 1973, \aap, 24, 337

\bibitem[{{Sirono}(1999)}]{1999A&A...347..720S}
{Sirono}, S. 1999, \aap, 347, 720

\bibitem[{{Tanaka} {et~al.}(2002){Tanaka}, {Takeuchi}, \&
  {Ward}}]{2002ApJ...565.1257T}
{Tanaka}, H., {Takeuchi}, T., \& {Ward}, W.~R. 2002, \apj, 565, 1257

\bibitem[{{Tazzari} {et~al.}(2017){Tazzari}, {Testi}, {Natta}, {Ansdell},
  {Carpenter}, {Guidi}, {Hogerheijde}, {Manara}, {Miotello}, {van der Marel},
  {van Dishoeck}, \& {Williams}}]{2017A&A...606A..88T}
{Tazzari}, M., {Testi}, L., {Natta}, A., {et~al.} 2017, \aap, 606, A88

\bibitem[{{Visser} \& {Ormel}(2016)}]{2016A&A...586A..66V}
{Visser}, R.~G. \& {Ormel}, C.~W. 2016, \aap, 586, A66

\bibitem[{{Wada} {et~al.}(2013){Wada}, {Tanaka}, {Okuzumi}, {Kobayashi},
  {Suyama}, {Kimura}, \& {Yamamoto}}]{2013A&A...559A..62W}
{Wada}, K., {Tanaka}, H., {Okuzumi}, S., {et~al.} 2013, \aap, 559, A62

\bibitem[{{Wada} {et~al.}(2009){Wada}, {Tanaka}, {Suyama}, {Kimura}, \&
  {Yamamoto}}]{2009ApJ...702.1490W}
{Wada}, K., {Tanaka}, H., {Suyama}, T., {Kimura}, H., \& {Yamamoto}, T. 2009,
  \apj, 702, 1490

\bibitem[{{Weidenschilling}(1977)}]{Weidenschilling1977}
{Weidenschilling}, S.~J. 1977, \mnras, 180, 57

\bibitem[{{Whipple}(1972)}]{Whipple1972}
{Whipple}, F.~L. 1972, in From Plasma to Planet, ed. A.~{Elvius}, 211

\bibitem[{{Yang} {et~al.}(2017){Yang}, {Johansen}, \&
  {Carrera}}]{2017A&A...606A..80Y}
{Yang}, C.-C., {Johansen}, A., \& {Carrera}, D. 2017, \aap, 606, A80

\bibitem[{{Yang} {et~al.}(2013){Yang}, {Ciesla}, \&
  {Alexander}}]{2013Icar..226..256Y}
{Yang}, L., {Ciesla}, F.~J., \& {Alexander}, C.~M.~O.~. 2013, \icarus, 226, 256

\bibitem[{{Youdin} \& {Goodman}(2005)}]{2005ApJ...620..459Y}
{Youdin}, A.~N. \& {Goodman}, J. 2005, \apj, 620, 459

\bibitem[{{Youdin} \& {Lithwick}(2007)}]{2007Icar..192..588Y}
{Youdin}, A.~N. \& {Lithwick}, Y. 2007, \icarus, 192, 588

\bibitem[{{Youdin} \& {Shu}(2002)}]{2002ApJ...580..494Y}
{Youdin}, A.~N. \& {Shu}, F.~H. 2002, \apj, 580, 494

\bibitem[{{Zsom} {et~al.}(2010){Zsom}, {Ormel}, {G{\"u}ttler}, {Blum}, \&
  {Dullemond}}]{2010A&A...513A..57Z}
{Zsom}, A., {Ormel}, C.~W., {G{\"u}ttler}, C., {Blum}, J., \& {Dullemond},
  C.~P. 2010, \aap, 513, A57

\end{thebibliography}

\begin{sidewaystable*}
\caption{Parameter values and results for different model runs.}
\label{tab:parstudy}
\centering
\begin{tabular}{l l l l l l l l l l | l l l}
\hline
\multicolumn{10}{c |}{\bf Model parameters} & \multicolumn{3}{c}{\bf Results}\\
& & & & & & & & & & & &\\
ID & $M_{\star}$ & $\dot{M}_{\rm{gas}}$ & $\alpha$ & $M_{\rm{disk}}$ & $r_{\rm{out}}$ & $T_{\rm{visc, 1 au}}$ & $T_{\rm{irr, 1 au}}$ & specialties & $M_{\rm{solids,start}}$ & $M_{\rm{pltsml,icy}}$ & $M_{\rm{pltsmls,dry}}$ & $t_{\rm{peak,start}}$ -- $t_{\rm{peak,end}}$ \\
 & $[M_{\odot}]$ & $[M_{\odot} \: \rm{yr}^{-1}]$ & [] & $[M_{\star}]$ & $[\rm{au}]$ & $[\rm{K}]$ & $[\rm{K}]$ & & $[M_{\oplus}]$ & $[M_{\oplus}]$ & $[M_{\oplus}]$ & [yr]\\
\hline
\hline
1a \bf{(fid)} & 1 & $5.0 \times 10^{-9}$ & $10^{-3}$ & 0.040 & 55 & 350 & 177 & & 107.5 & 9.7 & 0 & $9.9 \times 10^{2}$ -- $3.2 \times 10^{4}$\\
1b & 1 & $5.0 \times 10^{-9}$ & $2 \times 10^{-3}$ & 0.040 & 108 & 350 & 177 & & 108.2 & 0 & 0 & -\\
1c & 1 & $5.0 \times 10^{-9}$ & $5 \times 10^{-3}$ & 0.040 & 268 & 350 & 177 & & 109.1 & 0 & 0 & -\\
1d & 1 & $5.0 \times 10^{-9}$ & $2 \times 10^{-3}$ & 0.056 & 150 & 350 & 177 & & 151.4 & 0 & 0 & -\\
1e & 1 & $5.0 \times 10^{-9}$ & $5 \times 10^{-3}$ & 0.022 & 150 & 350 & 177 & & 60.6 & 0 & 0 & -\\
1f & 1 & $5.0 \times 10^{-9}$ & $10^{-2}$ & 0.011 & 150 & 350 & 177 & & 30.3 & 0 & 0 & -\\
1g & 1 & * & $3 \times 10^{-2}$ & 0.040 & 161 & * & 177 & * $\dot{M}_{\rm{gas}} = f(t)$, $v_{\rm{frag, ice}} = 60 \: \rm{m} \: \rm{s}^{-1}$ & 104.5 & 0 & 0 & -\\
1h & 1 & $5.0 \times 10^{-9}$ & $10^{-3}$ & 0.040 & 55 & 350 & 177 & including pebble accretion & 107.5 & 89.9 & 0 & $9.9 \times 10^{2}$ -- $7.9 \times 10^{3}$\\
1i & 1 & $5.0 \times 10^{-9}$ & $10^{-3}$ & 0.040 & 55 & 350 & 177 & $Z = 0.02$ & 215.0 & 103.6 & 4.6 & $4.2 \times 10^{2}$ -- $4.8 \times 10^{4}$\\
1j & 1 & $5.0 \times 10^{-9}$ & $10^{-3}$ & 0.040 & 55 & 350 & 177 & $Z = 0.02$, $\epsilon_{\rm{SI}}$ = 0.1 & 215.0 & 103.6 & 8.9 & $4.2 \times 10^{2}$ -- $4.8 \times 10^{4}$\\
1k & 1 & $5.0 \times 10^{-9}$ & $10^{-3}$ & 0.040 & 55 & 350 & 177 & $Z = 0.02$, $\epsilon_{\rm{SI}}$ = 1.0 & 215.0 & 103.6 & 12.6 & $4.2 \times 10^{2}$ -- $4.8 \times 10^{4}$\\
1l & 1 & $5.0 \times 10^{-9}$ & $10^{-3}$ & 0.040 & 55 & 350 & 177 & $v_{\rm{frag, ice}} = 60 \: \rm{m} \: \rm{s}^{-1}$ & 107.5 & 17.1 & $1 \times 10^{-7}$ & $1.2 \times 10^{3}$ -- $2.2 \times 10^{4}$\\
\hline
2a & 0.1 & $3.2 \times 10^{-11}$ & $10^{-4}$ & 0.040 & 31 & 55 & 18 & & 10.3 & 4.6 & $8 \times 10^{-3}$ & $1.2 \times 10^{2}$ -- $6.9 \times 10^{4}$\\
2b & 0.1 & $3.2 \times 10^{-11}$ & $10^{-3}$ & 0.040 & 268 & 55 & 18 & & 10.7 & 1.0 & $2 \times 10^{-5}$ & $1.5 \times 10^{2}$ -- $3.7 \times 10^{5}$\\
2c & 0.1 & $3.2 \times 10^{-11}$ & $2 \times 10^{-3}$ & 0.020 & 268 & 55 & 18 & & 5.3 & 0.2 & 0 & $1.8 \times 10^{3}$ -- $1.5 \times 10^{5}$\\
2d & 0.1 & $3.2 \times 10^{-11}$ & $10^{-3}$ & 0.022 & 150 & 55 & 18 & & 5.9 & 0.8 & $2 \times 10^{-5}$ & $1.5 \times 10^{2}$ -- $2.3 \times 10^{5}$\\
2e & 0.1 & $3.2 \times 10^{-11}$ & $2 \times 10^{-3}$ & 0.011 & 150 & 55 & 18 & & 2.9 & 0.2 & 0 & $1.8 \times 10^{3}$ -- $1.3 \times 10^{5}$\\
\hline
3a & 0.5 & $1.1 \times 10^{-9}$ & $10^{-3}$ & 0.040 & 88 & 201 & 89 & & 53.7 & 5.8 & $4 \times 10^{-2}$ & $5.4 \times 10^{2}$ -- $6.1 \times 10^{4}$\\
3b & 0.5 & $1.1 \times 10^{-9}$ & $2 \times 10^{-3}$ & 0.040 & 174 & 201 & 89 & & 54.2 & 0.1 & 0 & $1.1 \times 10^{4}$ -- $2.8 \times 10^{4}$\\
3c & 0.5 & $1.1 \times 10^{-9}$ & $5 \times 10^{-3}$ & 0.040 & 429 & 201 & 89 & & 54.1 & 0 & 0 & -\\
3d & 0.5 & $1.1 \times 10^{-9}$ & $10^{-3}$ & 0.069 & 150 & 201 & 89 & & 93.1 & 6.7 & $5 \times 10^{-2}$ & $5.4 \times 10^{2}$ -- $8.4 \times 10^{5}$\\
3e & 0.5 & $1.1 \times 10^{-9}$ & $2 \times 10^{-3}$ & 0.034 & 150 & 201 & 89 & & 46.6 & 0.1 & 0 & $1.1 \times 10^{4}$ -- $2.8 \times 10^{4}$\\
3f & 0.5 & $1.1 \times 10^{-9}$ & $5 \times 10^{-3}$ & 0.014 & 150 & 201 & 89 & & 18.6 & 0 & 0 & -\\
\hline
4a & 2 & $2.3 \times 10^{-8}$ & $10^{-3}$ & 0.040 & 34 & 610 & 354 & & 213.7 & 0 & 0 & -\\
4b & 2 & $2.3 \times 10^{-8}$ & $2 \times 10^{-3}$ & 0.040 & 67 & 610 & 354 & & 217.1 & 0 & 0 & -\\
4c & 2 & $2.3 \times 10^{-8}$ & $5 \times 10^{-3}$ & 0.040 & 165 & 610 & 354 & & 217.9 & 0 & 0 & -\\
4d & 2 & $2.3 \times 10^{-8}$ & $10^{-2}$ & 0.040 & 328 & 610 & 354 & & 218.0 & 0 & 0 & -\\
4e & 2 & $2.3 \times 10^{-8}$ & $2 \times 10^{-3}$ & 0.091 & 150 & 610 & 354 & & 494.6 & 0 & 0 & -\\
4f & 2 & $2.3 \times 10^{-8}$ & $5 \times 10^{-3}$ & 0.036 & 150 & 610 & 354 & & 197.9 & 0 & 0 & -\\
4g & 2 & $2.3 \times 10^{-8}$ & $10^{-2}$ & 0.018 & 150 & 610 & 354 & & 98.9 & 0 & 0 & -\\
\hline
\noalign{\vskip 3mm}   
\multicolumn{13}{l}{{\bf Notes.} From left to right, column entries denote: model ID number; stellar mass; gas accretion rate; turbulence strength parameter $\alpha$; gas disk mass; outer disk radius;}\\
\multicolumn{13}{l}{viscous temperature at 1 au; irradiated temperature at 1 au; model specialties; initial mass in solids; final mass in icy planetesimals; final mass in dry planetesimals;}\\
\multicolumn{13}{l}{time period during which streaming instability outside the snowline is operational (a `-' means no streaming instability outside the snowline has occurred).}\\
\end{tabular}
\end{sidewaystable*}

\appendix
\section{Validation}\label{sec:validation}
We test our Lagrangian smooth-particle model of dust growth and radial drift against analytical results from the literature.
\begin{figure}[t]
	\centering
		\includegraphics[width=0.5\textwidth]{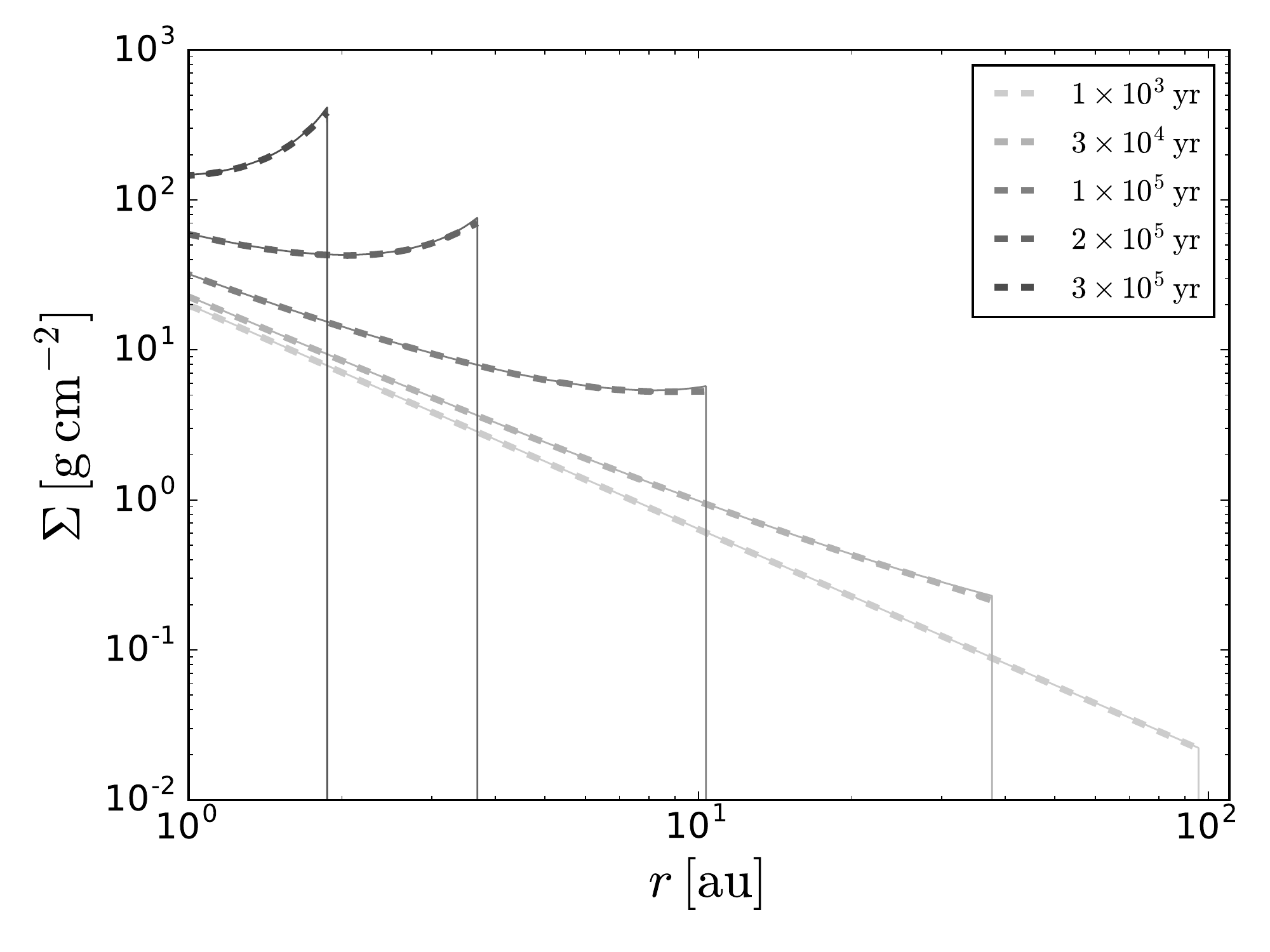}
\caption{Solids surface density $\Sigma$ as function of radial distance from the star $r$, plotted at different time points. Solid lines correspond to the analytical predictions from \citet{2002ApJ...580..494Y} (\eq{YS}), and scattered points correspond to our model results. The physical size of particles is 0.5 cm.\label{fig:YS}}
\end{figure}

\subsection{Drift-only}
First, we check if our model reproduces the analytical predictions of \citet{2002ApJ...580..494Y} for the surface density evolution of a single particle size (0.5 cm) without growth. 
We take a gas disk with surface density profile $\Sigma_{\rm{gas}} = 1000 \: (r / 1 \: \rm{au})^{-3/2} \: \rm{g} \: \rm{cm}^{-2}$ and temperature profile $T = 150 \: (r / 3 \: \rm{au})^{-1/2} \: K$.
We consider an initial solids surface density profile $\Sigma (r, 0)$ that has a cutoff at outer radius $r_{\rm{out}}: \Sigma (r, 0) = \Sigma_{1} (r / r_{1})^{-3/2}$ with $r_{1}$ = 1 au if $r < r_{\rm{out}}$ and $\Sigma (r, 0) = 0$ otherwise. Following the derivation in Sect.~4.1 of \citet{2002ApJ...580..494Y}, we then find that $\Sigma$ evolves with time $t$ as:
\begin{equation}\label{eq:YS}
\Sigma (r, t) = \Sigma_{1} r^{-d-1} r_{\rm{i}}^{d} (r, t), 
\end{equation}
where $d$ is defined through the radial dependence of the drift velocity $v_{\rm{dr}}$ (the expression for which we copy from \citet{2002ApJ...580..494Y}) for a constant particle size: $v_{\rm{dr}} \propto r^{d}$. For our temperature profile $T \propto r^{-0.5}$ and gas surface density profile $\Sigma_{\rm{gas}} \propto r^{-3/2}$ we find $d = 3/2$. $r_{\rm{i}} (r, t)$ is the initial location of a particle that ends up at radius $r$ at time $t$, which is given by:
\begin{equation}
r_{\rm{i}} (r, t) = r \left[ 1 - (d - 1) \frac{v_{\rm{dr}} (r) t}{r}\right]^{-\frac{1}{d-1}}.
\end{equation}
In \fg{YS} we compare \eq{YS} with our numerical results at different points in time, demonstrating that our model reproduces the analytical results presented by \citet{2002ApJ...580..494Y}.

\subsection{Including simple growth}
We next turn to \citet{2014A&A...572A.107L} (hereafter: LJ14) to compare the results of our model including particle growth with their analytical results. We take the same disk model as LJ14 and plot the resulting surface densities at different points in time in \fg{LJ1}. LJ14 introduced the term `pebble front' for the distance from the star up to which particles have grown to pebble-sizes and are drifting inwards. This distance, calculated by their Eq.~10, is plotted by the vertical dashed lines in \fg{LJ1}. Rather than a razor-sharp transition between the pebble region to the dust region as in the analytical work of LJ14, our numerical results show a smooth connection between the two regions, the radius of which is in agreement with LJ14. The pebble front radius that follows from our simulation progresses a little bit faster in time than the pebble front radius in LJ14, which is due to the fact that in their work, a constant value of $\xi$ (the number of growth e-foldings needed to grow to pebble-sizes), was used regardless of semi-major axis, whereas in our model it is not. The slope of the surface density interior to the pebble front falls off as $r^{-3/4}$, as was also found by LJ14.

\begin{figure}[t]
	\centering
		\includegraphics[width=0.49\textwidth]{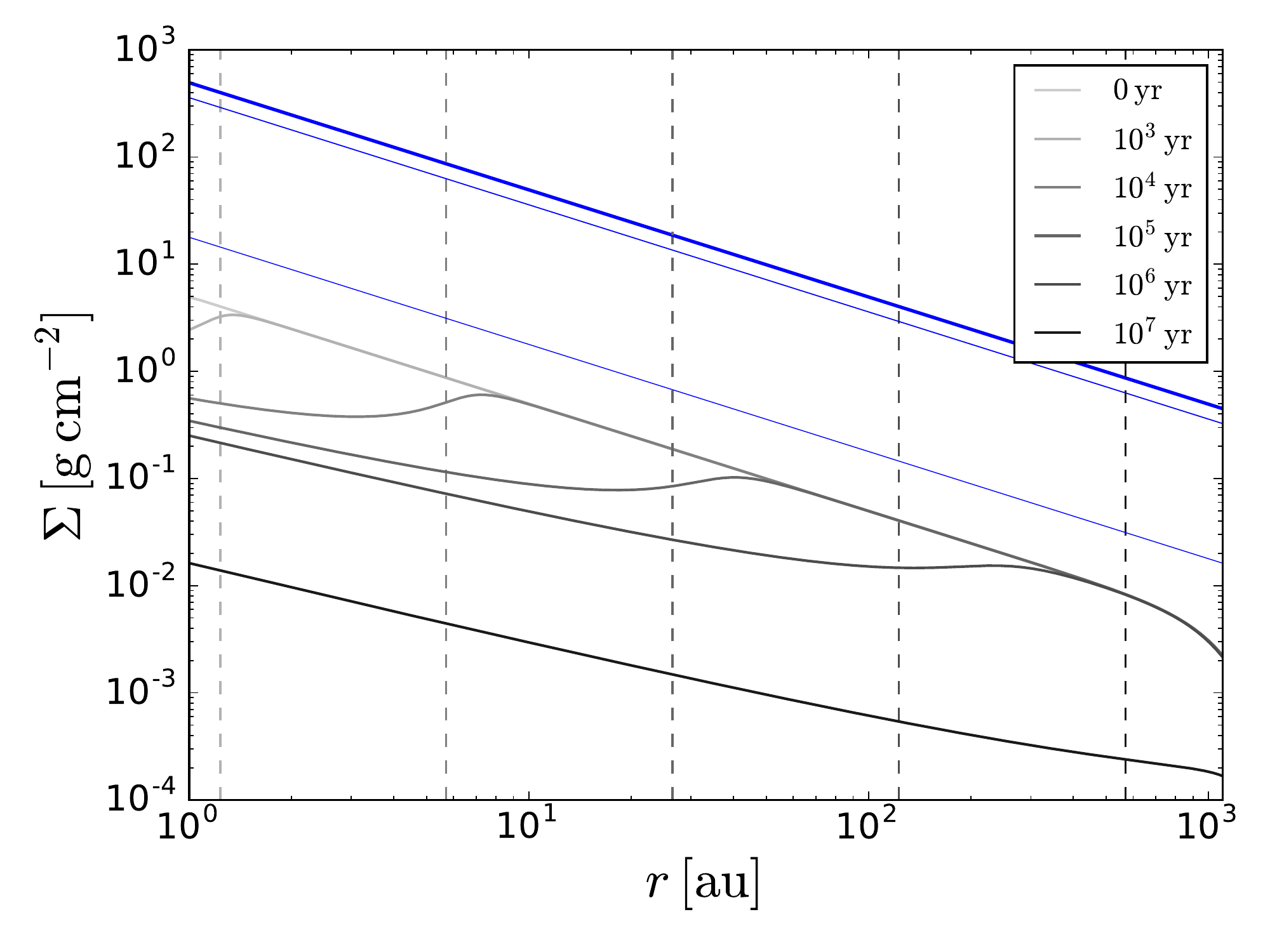}
\caption{Reproduction of Fig.~1 of \citet{2014A&A...572A.107L} with our model. Grey lines correspond to solids surface densities at different points in time (labels denote time in years). Gas surface density profiles are plotted in blue for the same time points. The vertical dashed lines indicate the radial extent of the pebble front defined in Eq.~10 of \citet{2014A&A...572A.107L}. \label{fig:LJ1}}
\end{figure}

\section{Updated semi-analytic model for solids enhancement outside the snowline}
Appendix~C of \citet{SO2017} contains several minor errors or unclarified approximations:
\begin{itemize}
    \item The term $\mathcal{M}_\mathrm{tot,ice}$ was omitted in the top line of Eq.~(C.10);
    \item In Eq.~(C.10) the normalized $\mathcal{M}_\mathrm{eq}$ of Eq.(C.7) was substituted. However, the difference between the particle and gas diffusivities was not accounted for;
    \item Eq.~(C.11) also implicitly assumes $D_\mathrm{gas}=D_p$.
\end{itemize}
As a result the semi-analytic model will give an erroneous result for particles reaching dimensionless stopping time approaching unity (at the location of the ice peak), since in that case $D_p<D_\mathrm{gas}$. For particles $\tau\ll1$, however, the model outlined in SO17 remains correct.

In this Appendix, we briefly re-derive the key expressions. 

Starting from Eq.~(C.8) of SO17, we normalize lengths to $r_\mathrm{snow}$, surface density to $\Sigma_\mathrm{snow}$ and velocities to $D_\mathrm{gas}/r_\mathrm{snow}$:
\begin{equation}
    \tilde\Sigma \tilde{v}_\mathrm{peb} +\frac{D_p}{D_\mathrm{gas}} \tilde\Sigma' = \frac{\mathcal{M}_z - \mathcal{M}_\mathrm{tot,ice}}{\Sigma_\mathrm{snow} D_\mathrm{gas}/r_\mathrm{ice}}.
\end{equation}
The left-hand side represents the mass flux of the ice; $\sim$ denote non-dimensional quantities. The ice surface density ($\Sigma$) is normalized by the quantity $\Sigma_\mathrm{snow}$ -- the surface density in H$_2$O vapor at the snowline $r_\mathrm{snow}$ -- the pebble drift velocity is normalized by $D_\mathrm{gas}/r_\mathrm{snow}$ and the radius $r$ is normalized by $r_\mathrm{snow}$.  The first term on the right-hand side (RHS) is given by Eq.(C.7) of SO17. The second term on the RHS is equals $+b_\mathrm{gas}=3/2$ (because $\mathcal{M}_\mathrm{tot,ice}$, the total mass flux in ice, is directed inwards) in an $\alpha$-disk. Applying these, we obtain:
\begin{equation}
    \tilde\Sigma \tilde{v}_\mathrm{peb} +\frac{D_p}{D_\mathrm{gas}} \tilde\Sigma' 
    = (a_\mathrm{eq} -b_\mathrm{gas}) e^{-a_\mathrm{eq}x} +b_\mathrm{gas}
\end{equation}
where $x=r/r_\mathrm{snow}-1$ as in SO17 and $a_\mathrm{eq}$ is a constant that enters the expression for the equilibrium vapor density of H$_2$O beyond the snowline.  Getting rid of $\sim$ notation and putting $D_\mathrm{gas}=r_\mathrm{snow}=1$:
\begin{equation}
    \label{eq:new-diff-eq}
    \Sigma {v}_\mathrm{peb} +D_p \Sigma' 
    = (a_\mathrm{eq} -b_\mathrm{gas}) e^{-a_\mathrm{eq}x} +b_\mathrm{gas}
\end{equation}
An analytical solution to this differential equation exists:
\begin{equation}
    \Sigma(x) = \frac{a_\mathrm{eq}b_\mathrm{gas}D_p(1-e^{-x v_\mathrm{peb}/D_p}) 
    +\left( a_\mathrm{eq}e^{-x v_\mathrm{peb}/D_p} +(b_\mathrm{gas}-a_\mathrm{eq})e^{-a_\mathrm{eq}x} -b_\mathrm{gas} \right)v_\mathrm{peb} }
    {(a_\mathrm{eq} D_p-v_\mathrm{peb})v_\mathrm{peb}}.
\end{equation}
Of particular importance is the location where this function peaks ($\Sigma'=0$). The maximum corresponding to $\Sigma(x)$ is located at
\begin{equation}
    x_\mathrm{peak} = \frac{D_p}{a_\mathrm{eq}D_p -v_\mathrm{peb}} \log \left[ \frac{(a_\mathrm{eq}-b_\mathrm{gas})D_p}{v_\mathrm{peb}-b_\mathrm{gas}D_p} \right] 
\end{equation}
and the corresponding value of the (normalized) surface density is
\begin{equation}
    \displaystyle
    \Sigma_\mathrm{peak}
    \displaystyle
    = \frac{1}{v_\mathrm{peb}} \left[ b_\mathrm{gas} +(a_\mathrm{eq}-b_\mathrm{gas}) e^{-a_\mathrm{eq} x_\mathrm{peak}}\right] 
\end{equation}
which follows from \eq{new-diff-eq}.

\end{document}